\documentclass[pra,twocolumn,superscriptaddress,showpacs,amsmath,amstex,amssymb,citeautoscript]{revtex4-1}

\usepackage[T1]{fontenc}
\usepackage[utf8]{inputenc}
\usepackage{lipsum}
\usepackage{amsmath}
\usepackage{amssymb}
\usepackage{bbm}
\usepackage{braket}
\usepackage{xcolor}
\usepackage{pifont}
\usepackage[mathscr]{euscript}
\usepackage[shortlabels]{enumitem}
\usepackage[justification=justified]{subcaption}
\usepackage{graphicx}
\usepackage{lipsum}
\allowdisplaybreaks
\usepackage{float}
\usepackage{graphicx}
\usepackage{dsfont}
\usepackage{comment}
\usepackage[colorlinks=true]{hyperref}  
\hypersetup{
    bookmarks=true,         % show bookmarks bar?
    unicode=false,          % non-Latin characters 
    pdftoolbar=true,        % show Acrobat
    pdfmenubar=true,        % show Acrobat 
    pdffitwindow=false,     % window fit to page when opened
    pdfstartview={FitH},    % fits the width of the page to the window
    pdftitle={Noise spectroscopy of insulating and itinerant altermagnets},    % title
    pdfauthor={Pupim and Scheurer},     % author
    pdfsubject={},   % subject of the document
    pdfcreator={},   % creator of the document
    pdfproducer={}, % producer of the document
    pdfkeywords={} {} {}, % list of keywords
    pdfnewwindow=true,      % links in new window
    colorlinks=true,       % false: boxed links; true: colored links
    linkcolor=blue, %red,          % color of internal links (change box color with linkbordercolor)
    citecolor=blue,        % color of links to bibliography
    filecolor=magenta,      % color of file links
    urlcolor=blue           % color of external links
}

\newcommand{\equref}[1]{Eq.~(\ref{#1})}
\newcommand{\equsref}[2]{Eqs.~(\ref{#1}) and (\ref{#2})}
\newcommand{\secref}[1]{Sec.~\ref{#1}}
\newcommand{\figref}[1]{Fig.~\ref{#1}}
\newcommand{\refcite}[1]{Ref.~\onlinecite{#1}}

\newcommand{\appref}[1]{Appendix~\ref{#1}}

\newcommand{\pdagger}{{\phantom{\dagger}}}
\newcommand{\diff}{\mathrm{d}}

\renewcommand{\approx}{\simeq}

\renewcommand{\Im}{\text{Im}}

\renewcommand{\vec}[1]{\boldsymbol{#1}}

\definecolor{wrongultramarine}{rgb}{1,0.5,0}
\begin{document}

\title{Noise spectroscopy of insulating and itinerant altermagnets}

\author{Lucas V. Pupim}
\affiliation{Institute for Theoretical Physics III, University of Stuttgart, 70550 Stuttgart, Germany}

\author{Mathias S.~Scheurer}
\affiliation{Institute for Theoretical Physics III, University of Stuttgart, 70550 Stuttgart, Germany}

\begin{abstract}
One of the central goals in the emergent field of altermagnetism is the unambiguous experimental identification and characterization of altermagnetic order across a variety of compounds. This motivates exploring tools that can clearly distinguish altermagnets from antiferromagnets, based on symmetry signatures, and offer access to the dominant orbital character (e.g., $d$-wave vs.~$g$-wave) of the magnetic order parameter. 
In this work, we theoretically explore the potential of noise magnetometry for this task, studying contributions from both magnons and itinerant electrons in different regimes and scenarios. While altermagnetism and antiferromagnetism also lead to different noise spectra for magnons, we find the most striking and symmetry-sensitive signatures in the charge fluctuations of itinerant altermagnets. Both for the homogeneous bulk case and in the presence of strain and/or around domain walls, we identify noise contributions that are only permitted by symmetry in the altermagnet and, thus, provide a unique signature of altermagnetism. Furthermore, the angular dependence of noise around domain walls also offers access to the orbital character of the altermagnet. On a more technical note, we discuss the role and relevance of lattice effects related to the dipole tensor. We hope that our work will help pave the way towards the clear experimental identification of altermagnetism across a wide range of candidate materials.
\end{abstract}

%\date{\today}
\maketitle
%\tableofcontents

\section{Introduction}
While noise is typically regarded as a complication for quantum computation that has to be suppressed, the sensitivity of qubits to noise can also be seen as an opportunity to probe other systems via noise spectroscopy~\cite{rovny2024nanoscale,gong2026spinrelaxometrysolidstatedefects,RevModPhys.89.035002,schoelkopf2003qubits}. Common platforms for such noise measurements are  Nitrogen vacancies in diamonds (NV centers)  \cite{taylor2008high,myers2014probing,romach2015spectroscopy,kolkowitz2015probing} and superconducting qubits \cite{bylander2011noise}. Using Pauli matrices $\vec{\sigma} = (\sigma_x,\sigma_y,\sigma_z)^T$ to describe the two-level system of the qubit, its bare Hamiltonian reads as $\mathcal{H}_{\mathrm{q}} = \omega_{\text{q}} \hat{\vec{n}}_{\text{q}}\cdot\vec{\sigma}/2$, where the unit vector $\hat{\vec{n}}_{\text{q}}$ is the qubit quantization axis and $\omega_{\text{q}}$ the qubit frequency.

In the spatial proximity of another system, charge and/or spin fluctuations in it lead to a time-dependent magnetic field $\vec{B}(t)$ at the qubit position, described by the additional contribution $\mathcal{H}_{\mathrm{q}-\mathrm{m}}=\mu_{\mathrm{B}} \boldsymbol{\sigma} \cdot \boldsymbol{B}(t)$ to the qubit Hamiltonian, which induces both relaxation and dephasing. This has been explored---both theoretically and experimentally---as a means to probe a variety of interesting many-body phenomena, such as quantum magnetism  \cite{van2015nanometre,PhysRevLett.121.187204,finco2021imaging,PhysRevB.98.180409,PhysRevB.99.104425,ziffer2024quantum,doi:10.1126/sciadv.adu9381,2025arXiv250119165X,PhysRevB.107.195102,PhysRevB.106.115108}
and its phase transitions \cite{PhysRevLett.131.070801}, broken time-reversal symmetry \cite{9h4l-21mt}, superconducting phases \cite{PhysRevB.105.024507,PhysRevResearch.4.L012001,1fzm-pb1d,liu2025quantumnoisespectroscopysuperconducting}, 1D channels \cite{PhysRevB.98.195433}, Dirac fermions \cite{2025PhRvB.111g5406M}, and correlated 2D metals \cite{PhysRevB.95.155107}.

The key information about the noise is encoded in the magnetic noise tensor defined as the correlator
\begin{equation}
    \mathcal{N}_{ab}(\omega) =\frac{1}{2} \int dt e^{i \omega t} \langle\{B_a(t),B_b(0) \} \rangle. \label{DefinitionOfCorrelator}
\end{equation}
Using Fermi's golden rule (see, e.g., \cite{PhysRevB.98.195433}), one can relate the relaxation rate of the qubit to the transverse components (perpendicular to $\hat{\vec{n}}_{\text{q}}$) of the noise tensor at the qubit frequency $\omega_q$, i.e., $1/T_1 \propto \mathcal{N}_{\perp}(\omega_{\text{q}})$. The dephasing rate, in contrast, is proportional to the longitudinal noise component $1/T_2 \propto \mathcal{N}_\parallel(\omega_{\mathrm{DD}})$ at the relevant frequency $\omega_{\mathrm{DD}}$ defined by the chosen dynamical decoupling protocol \cite{PhysRevLett.82.2417,PhysRevB.77.174509,PhysRevLett.107.230501,RevModPhys.89.035002,PhysRevLett.107.170504}.

Given the success of noise spectroscopy in analyzing magnetic systems, we here study the expected signatures for altermagnets (AMs), an emerging class of magnetic orders \cite{jungwirth2026symmetry,vsmejkal2020crystal,liu2025altermagnetismsuperconductivityshorthistorical,PhysRevX.12.040501,jungwirth2025altermagnetism,PhysRevX.12.031042}. These phases are characterized by a vanishing net magnetization (as opposed to ferromagnets) which is protected to be zero not as a result of translations, like in antiferromagnets (AFMs), but due to a combination of time-reversal and a point-group operation such as rotations.

\begin{figure}[b]
    \centering
    \includegraphics[width=\linewidth]{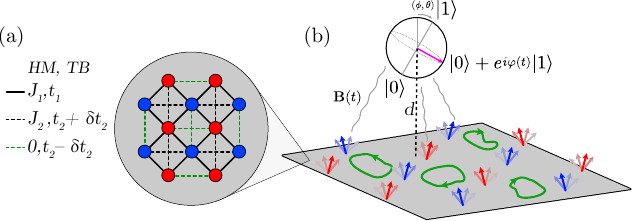}
    \caption{(a) Schematic checkerboard lattice and its couplings in the insulating (Heisenberg model) and metallic (tight-binding) regime. In the insulating regime, magnonic excitations generate a fluctuating magnetic field $\vec{B}(t)$, while in the conducting regime, current fluctuations cause these magnetic oscillations. The magnetic noise, in turn, is detected as shown in (b). The fluctuating magnetic field couples to the qubit and introduces noise into the coherence factor $\varphi(t)$. This noise, in turn, is detected as a decoherence rate $1/T_2$. Alternatively, one can start the qubit in the state $\ket{1}$ and see the effect of magnetic noise in the relaxation rate $1/T_1$.}
    \label{fig:1}
\end{figure}

To this end, we here consider models on the checkerboard lattice, \figref{fig:1}(a), which will allow us to stabilize both altermagnetic and antiferromagnetic phases, depending on model parameters, and thus to contrast them directly. We will first focus on the contributions to the noise coming from the fluctuations of the magnetic order parameter, which we will model using a Heisenberg spin model.

The spins $\vec{S}_j$ on the sites $j$ of the checkerboard lattice generate a magnetic field
\begin{equation}
\label{eq:B}
    \vec{B}=\frac{\mu_0 \mu_{\mathrm{B}}}{4 \pi} \sum_j\left[\frac{\vec{S}_j}{r_j^3}-\frac{3\left(\vec{S}_j \cdot \vec{r}_j\right) \vec{r}_j}{r_j^5}\right]
\end{equation}
in $\mathcal{H}_{\mathrm{q}-\mathrm{m}}$ due to the dipole-dipole interactions between the qubit and the spins. In \equref{eq:B}, $\vec{r}_j = (x_j,y_j,d)$ is the vector connecting the spin at lattice site $j$ and the qubit at a distance $d$ above the sample; see \figref{fig:1}(b) for a schematic illustration. Due to magnonic excitations above the long-range ordered ground state, these spins oscillate and produce magnetic fluctuations. As such, the noise will probe the magnon modes, which differ in AFMs and AMs as a result of their distinct symmetries. In particular, a well established fact is that AMs have split magnon bands \cite{PhysRevLett.133.156702,PhysRevLett.131.256703,bdfg-djbs}, which are degenerate in the antiferromagnetic limit. 

We focus on the regime where the magnon spectra are only partially gapped, due to an applied external magnetic field, and one band remains gapless. This means that the magnetic noise is dominated by single-magnon processes rather than two-magnon scattering and other higher processes. The contrasting ``diffusive regime'', where the spectrum is fully gapped (i.e., the probe frequency lies inside the magnon gap due to an easy-axis term) and single magnon processes do not contribute to the noise spectra was studied in \cite{bittencourt2025quantumimpuritysensingaltermagneticorder}.

While charge-carrying electronic excitations can be neglected in insulating AMs, such that magnons provide the key contributions to the noise spectrum, this is not necessarily the case in the metallic regime. In fact, one of the central properties of altermagnetic order is the resultant non-trivial splitting of the electronic bands and Fermi surfaces. As such, it is naturally expected to lead to non-trivial consequences for the noise arising from charge fluctuations.
To study this, we investigate an electronic tight-binding model on the same checkerboard lattice, \figref{fig:1}(a), with nearest and next-nearest neighbor hoppings. Now, charge fluctuations generating currents $\vec{J}_j(t)$ are the noise source and the magnetic field at the qubit position can be computed from the Biot-Savart law,
\begin{equation}
\vec{B} = \frac{\mu_0}{4\pi} \sum_{j} \frac{\vec{J}_j \times \vec{r}_j}{r_j^3}. \label{BiotSavart}
\end{equation}
More specifically, we look into how time-reversal symmetry breaking paired with other perturbations can give contrasting noise signatures for AFMs and AMs. This contrast, building on the approach of De \textit{et al.}~\cite{9h4l-21mt}, emerges in the relaxation rate difference between anti-aligned qubits $\delta \Gamma = 1/T_1(\hat{\vec{z}}) - 1/T_1(-\hat{\vec{z}})$. In the presence of strain or domain walls, only AMs show a finite $\delta \Gamma$, while AFMs do not produce any signal.

The remainder of the paper is organized as follows. In the first part, \secref{Insulating}, we will focus on insulating systems such that only fluctuations of the magnetic order parameter need to be taken into account. Specifically, in \secref{Model}, we present the spin model that we study in this work and discuss the resulting magnons. The computation of the resultant magnetic noise tensor can be found in \secref{MagneticNoise}, and \secref{Results} discusses the results for the relaxation rates. For concreteness, we restrict the presentation of explicit results to $1/T_2$, i.e., the longitudinal noise $\mathcal{N}_\parallel$, in the main text; the behavior of $1/T_1$ is shown in the appendix. Then we generalize to itinerant systems in \secref{Itinerant}. After defining the model we use (\secref{ModelItinerant}), we discuss the noise components induced by strain and around domain walls in \secref{Strain} and \secref{DomainWalls}, respectively. A symmetry analysis can be found in \secref{SymmetryConsiderations}. Our findings are summarized in \secref{Conclusion} and multiple appendices provide further details on the methodology, additional data, and symmetry arguments.

\section{Insulating altermagnets}\label{Insulating}

\subsection{Model and Magnons}\label{Model}
To describe altermagnetism and contrast it with antiferromagnetism in a minimal setting, we use a Heisenberg model on the checkerboard lattice, with Hamiltonian
\begin{equation}
    H = J_1 \sum_{\langle i,j\rangle} \vec{S}_i \cdot \vec{S}_j + J_2 \sum_{\langle\langle i,j\rangle\rangle} \vec{S}_i \cdot \vec{S}_j - \sum_i \vec{h}\cdot \vec{S}_i \,\,\,.
    \label{eq:Hamiltonian_1}
\end{equation}
The sum over $\langle i,j\rangle$ includes nearest neighbors (connecting different sublattices) and the sum over $\langle\langle i,j\rangle\rangle$ corresponds to the second nearest neighbors (connecting sites in the same sublattice) indicated as black dashed lines in \figref{fig:1}(a). We consider the distance between sites of the same sublattice to be $a=1$. In addition, we take the Zeeman field $\vec{h} =h \,\hat{\vec{z}}$ to be uniform and pointing out-of-plane. Notably, this model has been studied in the context of planar pyrochlore \cite{PhysRevB.69.214427,PhysRevB.85.205122,sadrzadeh2015phase,Li_2015,PhysRevB.67.054411,PhysRevB.65.184408,PhysRevB.68.144422} and layered titanades \cite{singh1998new}. More recently it has been used (as we do here) as a minimal model for altermagnets \cite{PhysRevB.110.144421,PhysRevResearch.7.023152,PhysRevLett.134.196701,bdfg-djbs,bittencourt2025quantumimpuritysensingaltermagneticorder,xg1x-sj4c,s31h-hk2v}.

We parameterize the coupling constants as $J_1 = \eta^2$ and $J_2 = -(1-\eta^2)$ for $ 0 \leq\eta\leq 1$, so that $|J_1|+|J_2| =1$ and consequently, the width of the magnon bands is independent of $\eta$. This will capture three different scenarios: when $\eta = 1$ (i.e., $J_1 =1, J_2 =0$), we effectively obtain a 2D (square-lattice) antiferromagnet with nearest neighbor exchange couplings; when $\eta = 0$ (i.e., $J_1 =0, J_2 =-1$), the system decouples into 1D (chains in the $x$ and $y$ directions) ferromagnets; finally, if $0 < \eta <1$, we have an altermagnet since, when both $J_1$ and $J_2$ are finite, the two sublattices in \figref{fig:1}(a) are not related by translation but, instead, by a rotation.
As the ferromagnetic chain limit is rather artificial and not of interest to us here, we here exclude this case, i.e. focus on $\eta \neq 0$. In fact, we will only consider $\eta \geq 1/\sqrt{2}$, to keep $|J_1| \geq |J_2|$. We note that the Zeeman field $\vec{h}$ also affects these orders by canting them out of the plane, the details of this effect will be discussed further on. 

%\textit{Magnons} -- 
As we are interested in the noise due to spin fluctuations, we consider the linear spin wave theory (LSWT) regime and calculate the magnon spectra of our model. To obtain these magnons we first perform a Holstein-Primakoff transformation \cite{PhysRev.58.1098}.
This can be done \cite{PhysRevLett.91.017205} by first rewriting the spin operators using a local basis as $\vec{S}_i = \vec{m}_i S^\parallel + \vec{S}^\perp_i$, where $\vec{m}_i$ is the direction of the local magnetization, $S^\parallel$ and $\vec{S}^\perp_i$ are the spin components parallel and perpendicular to $\vec{m}_i$, respectively. The perpendicular spin components are parameterized as $\vec{S}^{\perp}_{i} = \sum_{\nu=\pm} \vec{e}^s_{i}\,S^\nu_{i}$, with $\vec{e}^{\pm}_{i} = \tfrac{1}{2}\left( \vec{e}_{1,i} \mp i \vec{e}_{2,i} \right)$. Here, the unit vectors $\vec{e}_{1,i}$ and $\vec{e}_{2,i}$, together with $\vec{m}_i$, locally form a right-handed coordinate system in the sense that $\vec{m}_i = \left(\vec{e}_{1,i} \times \vec{e}_{2,i}\right)$.  

Now we use the Holstein-Primakoff transformation $\hat{S}^{\parallel}_{i} = S - \hat{b}^\dagger_{i}\hat{b}^\pdagger_{i}$ and $\hat{S}^+_{i} = (\hat{S}^-_{i})^\dagger 
\approx \sqrt{2S}\,\hat{b}_{i}$, where $\hat{b}_{i}$ are bosonic annihilation operators. We plug these expressions into \equref{eq:Hamiltonian_1} (for the detailed calculation, see \appref{app:h-p}) and find the direction of $\vec{m}_i$ by minimizing the classical energy. Writing $\vec{m}_i = \varrho_i \hat{\vec{x}} \cos{\theta_m}  + \hat{\vec{z}}\sin{\theta_m} $ with $\varrho_i = \pm1$ for different sublattices, this yields
\begin{equation}
\theta_{m} \;=\;
\begin{cases}
\arcsin\!\left(\dfrac{h}{8 J_1 S}\right), & |h| \leq 8 J_1 S, \\[10pt]
\dfrac{\pi}{2}, & |h| > 8 J_1 S .
\end{cases}
\end{equation}
Here we see explicitly the canting effect that the Zeeman field has, i.e., it tilts the compensated (in-plane) local magnetization out of the plane and generates an uncompensated component $\vec{m}_i \cdot \hat{\vec{z}} = \sin{\theta_m}$.

\begin{figure}[tb]
    \centering
    \includegraphics[width=\linewidth]{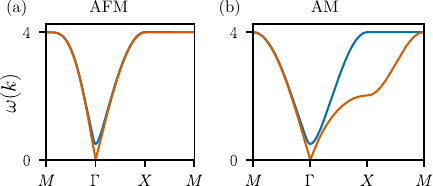}
    \caption{(a) Antiferromagnetic ($\eta=1$, i.e., $J_1 =1, J_2=0$) and (b) altermagnetic ($\eta=1/\sqrt{2}$, i.e, $J_1 = 0.5, J_2 = -0.5$) magnon dispersion for the model in \equref{eq:Hamiltonian_1}. The Zeeman field is $h=0.5$ for both cases.}
    \label{fig:2}
\end{figure}

Within LSWT the spin-model Hamiltonian in \equref{eq:Hamiltonian_1} is approximated by a quadratic bosonic theory, $H \rightarrow \mathcal{H}$, which can be compactly written in momentum ($\vec{k}$) space and in matrix form. We label operators $\hat{b}_{a}$ from different sublattices with the lower index $a=1,2$, and use the Bogoliubov basis $\hat{\psi}_{\vec{k}} = \left[\hat{b}_{1,{\vec{k}}}^\pdagger\,\, \hat{b}_{2,{\vec{k}}}^\pdagger \,\, \hat{b}^\dagger_{1,-{\vec{k}}} \,\, \hat{b}^\dagger_{2,-{\vec{k}}}\right]^T$, so $\mathcal{H}/S =\frac{1}{2} \sum_k \hat{\psi}_{\vec{k}}^\dagger\,\, h_{\vec{k}} \,\,\hat{\psi}^\pdagger_{\vec{k}}$,
\begin{widetext}
\begin{equation}
\label{eq:h_k}
   h_{\vec{k}} = \begin{pmatrix}
4 J_1 \cos{2\theta_m} + A_x& -C \,\sin^2{\theta_m} & 0 & C \cos^2{\theta_m}  \\
-C \,\sin^2{\theta_m} & 4 J_1 \cos{2\theta_m}  + A_y& C \cos^2{\theta_m} & 0\\
0&C \cos^2{\theta_m}&4 J_1 \cos{2\theta_m} + A_x & -C \,\sin^2{\theta_m}\\
C \cos^2{\theta_m}&0&-C \,\sin^2{\theta_m}& 4 J_1\cos{2\theta_m}  + A_y
\end{pmatrix}  +\frac{h}{ S} \sin{\theta_m} \mathbb{I}.
\end{equation}
\end{widetext}
For clarity in the expression, we suppressed the $\vec{k}$ dependence of the terms on the right-hand side and defined the momentum dependent factors $A_i = 2J_2(\cos{k_i}-1)$ and $C=4 J_1 \cos{(k_x/2)} \cos{(k_y/2)}$. We note that, without the Zeeman field, one could simplify the Hamiltonian to effectively become $2\times 2$ (akin to a superconductor with or without spin-orbit coupling) making it straightforward to diagonalize it analytically, c.f. \cite{PhysRevB.65.184408,bittencourt2025quantumimpuritysensingaltermagneticorder}.
However, in the ``full'' problem, there are no compact analytical expressions available. Hence, we diagonalize our quadratic bosonic Hamiltonian numerically through the Colpa method \cite{COLPA1978327,toth2015linear}. We give a concise summary of this method in \appref{app:colpa}.

In \figref{fig:2}, we can see the magnon modes for the antiferromagnetic (a) and altermagnetic (b) cases. The two phases share some features for small momentum, e.g., the linear Goldstone mode around $\Gamma$ and the gapped mode due to the Zeeman field. The substantial difference is seen closer to the edge of the Brillouin zone (around the $X$ point), where the two magnonic modes are degenerate for the AFM, while for the AM they are split; similar to the spin splitting of the electronic bands, this is a characteristic feature of magnons in altermagnets \cite{PhysRevLett.131.256703,PhysRevB.108.L100402}. %Of course, one could say that this is a defining characteristic of the AM, since it is a fundamental shared trait in candidate AM materials. 

Another fact that these AM materials and models have is the pair of Van Hove singularities pinned at the edge of the Brillouin zone. As we will see in more detail in the next section, this feature can be important for noise spectroscopy, since this method is not momentum resolved and primarily sensitive to the DOS.

\subsection{Computation of Noise Tensor}\label{MagneticNoise}

Now, we need to relate the spin waves from our Hamiltonian model to the magnetic noise sensed through the qubit dephasing (decoherence) or relaxation. For concreteness, we focus on the dephasing rate; however, a similar derivation follows for the relaxation rate, see \appref{app:noise}. Let us start with the qubit quantization axis $\vec{n}_q$ along the $z$ direction, i.e., $\vec{n}_q = (0,0,1)^T$, such that the associated rate is $1/T_2 \propto \mathcal{N}_{zz}$ with magnetic noise tensor components $\mathcal{N}_{ab}$ defined in \equref{DefinitionOfCorrelator}. We can rotate this axis to an arbitrary direction $\vec{n}_q = (\sin \theta \sin \phi,-\sin \theta \cos \phi,\cos \theta)^T$, see \figref{fig:1}(b), by a pair of rotations through the $z$ and $x$ axes $R_{\text{tot}} = R_z(\phi)R_x(\theta)$, such as $\mathcal{N} \rightarrow R_{\text{tot}}^T\, \mathcal{N}\, R_{\text{tot}} $. The relevant component becomes
\begin{equation}
\begin{split}
\mathcal{N}_{zz}\xrightarrow{R_{\text{tot}}} \,\, &\mathcal{N}_{zz}\cos^2\theta
+ \sin(2\theta)\,\sin\phi \,\,\mathcal{N}_{xz}\\
&+ \sin^2\theta \left( \mathcal{N}_{yy}\cos^2\phi + \mathcal{N}_{xx}\sin^2\phi \right).
\end{split}
\end{equation}
Each of these tensor components can, in turn, be written as functions of the spin-spin correlators, by inserting \equref{eq:B} into \equref{DefinitionOfCorrelator}. For conciseness, we here write down explicitly only $\mathcal{N}_{zz}$ and leave the detailed calculation for \appref{app:noise}, 
\begin{equation}
\label{eq:noise}
\begin{split}
    \mathcal{N}_{zz}  &=\frac{1}{2} \coth{\frac{\omega}{2T}} \left(\frac{\mu_0 \mu_{\mathrm{B}}}{4 \pi}\right)^2 \\ &\quad \times \int d^2\vec{k} \,\,  e^{-2kd} k^2 (\mathcal{C}_{zz} + \frac{\mathcal{C}_{xx} + \mathcal{C}_{yy}}{2}),\end{split}
\end{equation}
where $T$ is the temperature and
\begin{equation}
 \mathcal{C}_{ij}(\vec{k},\omega) = \sum_{a,a'} \Im\left[\,i\int_0^\infty dt e^{i\omega t} \langle [S_a^i(\vec{k},t),S_{a'}^j(-\vec{k},0)] \rangle\,\right].
\end{equation}
A well known property of the expression (\ref{eq:noise}), which also applies to the other components of the magnetic noise tensor, is the presence of a momentum filter $e^{-2kd}k^2$ that peaks at $k\sim 1/d$ while suppressing other values. One can picture this filter as the spatial averaging out of noise from short wavelength ($k\gg 1/d $) modes while long wavelength magnons ($k\ll 1/d $) are nearly uniform over the sensor scale and therefore generate little noise.

We also note that to calculate these spin-spin correlators, we need the spin operators in the lab frame, which read in terms of our Holstein-Primakoff bosons as 
\begin{equation}
\label{eq:spin-hp}
    \begin{split}
            S^x_a(\vec{k}) &\approx \varrho_a \left(\sqrt{2S} \sin \theta_m \frac{\hat{b}_{a,\vec{k}}^\pdagger + \hat{b}_{a,-\vec{k}}^\dagger}{2} \right. \\ &\qquad\qquad +\left. \cos \theta_m (S -\hat{b}_{a,\vec{k}}^\dagger  \hat{b}_{a,\vec{k}}^\pdagger ) \right) ,\\
         S^y_a(\vec{k}) &\approx i \,\varrho_a \sqrt{2S}\frac{\hat{b}_{a,\vec{k}}^\pdagger - \hat{b}_{a,-\vec{k}}^\dagger}{2} ,\\
         S^z_a(\vec{k}) &\approx - \varrho_a \sqrt{2S} \cos \theta_m \frac{\hat{b}_{a,\vec{k}}^\pdagger + \hat{b}_{a,-\vec{k}}^\dagger}{2}  \\ &\quad\,+\sin \theta_m (S -\hat{b}_{a,\vec{k}}^\dagger  \hat{b}_{a,\vec{k}}^\pdagger ),
    \end{split}
\end{equation}
where $a=1,2$ again labels the two sublattices.
We use the diagonalizing (Bogoliubov) transformation to rewrite our bosonic operators, i.e.
$\hat{\psi}_{\vec{k}} = T_{\vec{k}} \hat{\Psi}_{\vec{k}}$,
with
$\hat{\Psi}_{\vec{k}} =
\left(
\hat{\gamma}^{\pdagger}_{+,\vec{k}},
\hat{\gamma}^{\pdagger}_{-,\vec{k}},
\hat{\gamma}^{\dagger}_{+,-\vec{k}},
\hat{\gamma}^{\dagger}_{-,-\vec{k}}
\right)^T$ , which also allows to conveniently state their form in the Heisenberg picture, $\hat{\gamma}_{\pm, \vec k} (t) = e^{-i \omega_{\pm, \vec k } t } \hat{\gamma}_{\pm, \vec k} (0)$, where $\omega_{\pm, \vec k }$ are the associated magnon-band energies. The transformation $T_{\vec{k}}$ can be chosen to have all its entries being real numbers and we write
\begin{equation}
    T_{\vec{k}} = \begin{pmatrix}
        \mathcal{U}(\vec{k}) & \mathcal{V}(\vec{k})\\
        \mathcal{V}(\vec{k}) & \mathcal{U}(\vec{k})
    \end{pmatrix},
\end{equation}
where $\mathcal{U}$ and $\mathcal{V}$ are $2 \times 2$ matrices. Denoting their components by $u_{ij}= [\mathcal{U}(k)]_{ij}$ and $v_{ij}= [\mathcal{V}(k)]_{ij}$, the spin-spin correlators become
\begin{equation}
\label{eq:correlators}
\begin{split}
    &C_{xx} =\frac{S}{2} \sin^2\theta_m  \,\left[\delta(\omega - \omega_{+,\vec{k}}) (u_{11} - u_{12} + v_{11}- v_{21})^2 \right.\\
    &+\left. \delta(\omega - \omega_{-,\vec{k}}) (u_{12} - u_{22} + v_{12}- v_{22})^2  \right],\\
\\
    &C_{yy} =\frac{S}{2}\, \,\left[\delta(\omega - \omega_{+,\vec{k}}) (u_{11} - u_{12} - v_{11}+ v_{21})^2 \right.\\
    &+\left. \delta(\omega - \omega_{-,\vec{k}}) (u_{12} - u_{22} - v_{12}+ v_{22})^2  \right],\\
    \\
    &C_{zz} =\frac{S}{2} \cos^2\theta_m \,\left[\delta(\omega - \omega_{+,\vec{k}}) (u_{11} - u_{12} + v_{11}- v_{21})^2 \right.\\
    &+\left. \delta(\omega - \omega_{-,\vec{k}}) (u_{12} - u_{22} + v_{12}- v_{22})^2  \right],
    \end{split}
\end{equation}
where we suppressed the $\vec{k}$ dependence of the coherence factors $u_{ij}$ and $v_{ij}$ on the right-hand side for notational simplicity.
We note that these expressions are proportional to the spectral density of each band. This feature will give us qualitative differences between the AM and the AFM cases. We further note that the noise is dominated (in leading order) by single magnon processes due to our $(1/S)$ expansion. Hence, there is no contribution from the two-point correlator terms proportional to $
\sum_{\vec q}
\delta\!\left(\pm \omega_{\vec{k}+\vec{q}}
\pm \omega_{\vec{k}} - \omega\right)
\left(n_{\vec{k}\pm\vec{q}} \pm n_{\vec{k}}\right)$.

\subsection{Results}\label{Results}
In the expressions of \equref{eq:correlators}, we can see that the spin-spin correlators $C_{ij}(\vec{k},\omega)$ are all proportional to $\delta(\omega - \omega_{\pm,\vec{k}})$. Hence, the relevant components of $\mathcal{N}$, which involve integrals of $C_{ij}(\vec{k},\omega)$ over $\vec{k}$, are crucially determined by the density of states of the magnon bands $\omega_{\pm,\vec{k}}$. This fact will be the cause of the main qualitative difference in the noise spectrum between AFMs and AMs to be discussed next. 

Recall from \figref{fig:2}(a) that the antiferromagnetic magnon spectrum has a Van-Hove singularity only at the highest energy ($\omega_{\pm,\vec{k}} = 4$ in our dimensionless units) and pinned to the edges of the Brillouin zone. On the other hand, the altermagnetic magnon bands are split and therefore exhibit Van-Hove singularities at two different energies. For our specific model and choice of parameters ($\eta^2=1/2$) used in the numerics, the second Van-Hove energy is at half of the bandwidth ($\omega_{\pm,\vec{k}} = 2$ in our dimensionless units) and also pinned at the edge of the Brillouin zone. 

\begin{figure}[tb]
    \centering
    \includegraphics[width=\linewidth]{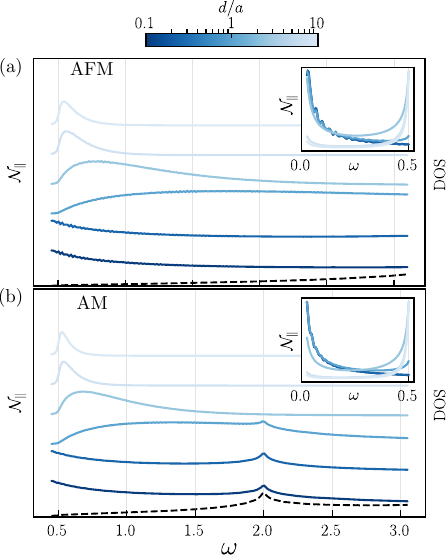}
    \caption{Parallel noise $\mathcal{N}_\parallel$ of a qubit probe aligned in-plane ($\theta=\pi/2,\phi=0$) for the (a) antiferromagnetic ($\eta=1$) and (b) altermagnetic ($\eta=1/\sqrt{2}$) cases. The black dashed line is the density of states (DOS, right axis). The rate curves are plotted for different distances (shade of blue) and they are shifted and rescaled for clarity. The insets show $\omega<h=0.5$. Here $T=1$.}
    \label{fig:3}
\end{figure}

This difference in the density of states is directly reflected in the decoherence rate $1/T_2$ when our probe qubit is very close to the sample, as shown in \figref{fig:3}. We can see that for both AFM and AM, the frequency $\omega$ dependence of $1/T_2$ is crucially determined by the DOS (dashed curves) behavior for $d<a$ (here we forego our convention of $a=1$ for clarity in the scales involved).
%However, as mentioned, the angular dependence of the noise does not affect the idea that the 
Consequently, for an AFM, $1/T_2(\omega)$ has a single peak in its decoherence (and also relaxation) profiles as one scans $\omega$, while altermagnetism leads to two peaks. We note that measuring $T_2$ (as opposed to $T_1$) allows us to scan $\omega$ by using different dynamic decoupling protocols \cite{PhysRevLett.82.2417,PhysRevB.77.174509,PhysRevLett.107.230501,RevModPhys.89.035002,PhysRevLett.107.170504} (frequency filters), e.g., the Hahn spin echo \cite{PhysRev.80.580}. 

Of course, the premise is that one can place the probe close enough to the sample and reach high frequencies. Unfortunately, this leads to practical issues that arise with this method of distinguishing an AFM and an AM. If one places the qubit farther and farther away from the magnetic sample, the different signatures of AFMs and AMs fade away. This problem comes from the term $e^{-2k d} k^2$ in the noise expression \equref{eq:noise}. As mentioned before, it acts as a momentum filter, favoring momenta with $k\sim 1/d$. This filter makes the effects that come from the edges of the Brillouin zone effectively invisible for distances $d\gg a$. Hence, the distinction between having one and two VHS in our model cannot be probed with qubits at distances larger than the lattice scale.
We note that this issue will become less severe in systems with Van-Hove singularities away from the edges of the Brillouin zone and, thus, at smaller momenta. Another possibility of resolving such complications lies in exploring systems where the lattice spacing $a$ is large enough, e.g., moiré altermagnets \cite{PhysRevLett.133.206702,2026arXiv260219734R}, so that $d \sim a$ becomes feasible. 

Note that, in the regime where $d\leq a$, the filter form is not precisely $e^{-2|\vec{k}| d} k^2$ since the Fourier transform of the magnetic dipole, \equref{eq:B}, needs to be performed on the lattice and the geometry of the sublattice needs to be taken into account. The noise expression for this situation is discussed in \appref{app:noise} but we emphasize that this is taken into account in all of our noise plots. A drastic contrast to the continuum limit can be seen in the insets of \figref{fig:3}. For small $d$ (darker curves), we find finite noise at $\omega =0$, while larger $d$ (lighter curves) has vanishing noise as a consequence of the momentum filtering.

We also note that for AFMs and AMs, the modes at the edge of the Brillouin zone typically have frequencies on the order of THz \cite{PhysRevLett.133.156702,PhysRevLett.131.256703,bdfg-djbs}, which is currently beyond what can be probed with available qubits and dynamic decoupling. As such, systems with smaller magnonic bandwidth are more naturally probed with noise spectroscopy, given the current status of the field. In a moiré system, this issue might be resolved due to the rescaling of effective interactions.   

\begin{figure}[tb]
    \centering
    \includegraphics[width=\linewidth]{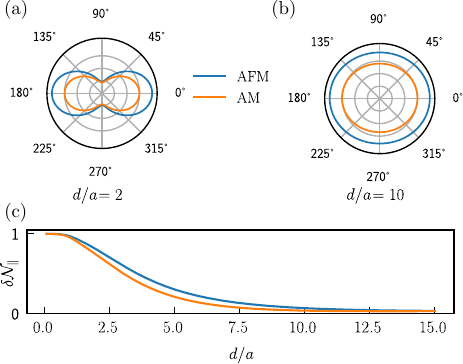}
    \caption{Angular dependence of parallel noise $\mathcal{N}_\parallel(\omega) \propto 1/T_2$. The qubit axis lies in the sample's plane ($\theta=\pi/2$) and we vary the azimuthal angle ($\phi$). In (a) we use $d=2$, while in (b) we use $d=10$. For larger distances ($d\gg a$), we note that the noise tends to becoming isotropic, while for shorter distances ($d \sim a$), $\mathcal{N}_\parallel(\omega)$ is suppressed when the qubit is aligned with the magnetic order ($\phi = 0^\circ, 180^\circ$). In (c) we show that the relative anisotropy $\delta \mathcal{N}_\parallel$ decays as $d$ increases, eventually showing the continuum limit, where the noise is isotropic. For all plots, we use $T=1$, $\omega=0.25$ and $h=0.5$. }
    \label{fig:angular}
\end{figure}

% Coherence factors and directional dependence
Apart from the density of states coming from the factors $\delta(\omega - \omega_{\pm,\vec{k}})$, one could expect that the Bogoliubov coefficients $u_{ij}(\vec{k})$ and $v_{ij}(\vec{k})$ [which are also present in the noise expression, see \equref{eq:correlators}] would generate further differences in noise profiles for the AFM and AM. However, we do not observe a qualitative difference between the two phases that is clearly rooted in these coherence factors. We can ascribe this lack of difference to the qubit ``feeling'' the noise isotopically. Mathematically, this fact emerges in the form of the momentum integral over the whole Brillouin zone in \equref{eq:noise}, which averages over all directions in momentum space.

If one calculates the relaxation rate $1/T_1$, which is sensitive to the transverse noise $\mathcal{N}_\perp$, qualitatively similar behavior follows. We can also rotate the qubit in any direction and obtain similar results. For instance, as one can see in \figref{fig:angular}(a), there is more noise when the qubit is perpendicular to the direction of the magnetic order (in our case, the $xz$ plane, i.e., $\phi = 0$ or $\pi$ in the angular plot) for short distances. As discussed previously, in this situation the sublattice information is still ``retained'' and we cannot use a continuum Fourier transform for the dipolar field \equref{eq:B}. In contrast, for \figref{fig:angular}(b), $\mathcal{N_\parallel}(\omega)$ shows an almost isotropic angular profile. As we further increase the distance of the qubit, we approach the regime where the continuum Fourier transform is valid and the noise is fully isotropic. We can explicitly see that by looking at the relative anisotropy $\delta \mathcal{N}_{\parallel} = |\mathcal{N}_{\parallel}(\phi=0^\circ) - \mathcal{N}_{\parallel}(\phi=90^\circ)|/|\mathcal{N}_{\parallel}(\phi=0^\circ) + \mathcal{N}_{\parallel}(\phi=90^\circ)|$ shown in \figref{fig:angular}(c).

Hence, the qubit alignment can make the noise stronger but does not provide extra \textit{qualitative} features that could serve as a smoking gun signature of the difference between AFM and AM. Of course, for a given model, like we study here, there are \textit{quantitative} differences between the two phases as is clearly visible in \figref{fig:angular}. As such, comparison of noise measurements with computations for specific models appropriate to a given altermagnetic candidate compound can indeed provide useful information about the underlying magnetic order.

\section{Noise from itinerant electrons}\label{Itinerant}
Having scrutinized the noise contributions from the magnetic order parameter in insulating AMs, we next generalize to itinerant systems. Here, the additional noise contributions from the gapless electronic excitations need to be taken into account. This is the focus of this section.

\subsection{Model}\label{ModelItinerant}
To describe itinerant electrons, we replace the Heisenberg model with an electronic tight-binding model on the checkerboard lattice in \figref{fig:1}(a). Instead of only spin degrees of freedom, electrons, as described by the creation operators $c^\dagger_{i,\sigma}$, $\sigma=\uparrow,\downarrow$, can now occupy the different sites $i = (\vec{R},a)$, labeled by the Bravais lattice points $\vec{R}$ and the sublattice index $a=1,2$. We include nearest-neighbor hopping (with amplitude $t_1$) along the solid diagonal bonds in \figref{fig:1}(a) as well as next-nearest-neighbor hopping; the associated amplitudes are given by $t_2 + \delta t_2$ and $t_2 - \delta t_2$ along the green and black dashed bonds, respectively. In the limit $\delta t_2 \rightarrow 0$, the checkerboard lattice becomes a square lattice with nearest- and next-nearest-neighbor hopping. To add antisymmetric spin-orbit coupling, we impose the $C_{4v}$ point-group symmetries and restrict ourselves to nearest-neighbor processes, such that spin-orbit coupling is described by a single parameter $\alpha_R$.

Introducing the associated momentum-space operators $c^\dagger_{\vec{k},a,\sigma}$, the Hamiltonian can be written as $H_{\text{it}} = H_{0} + H_{\text{m}}$, where $H_{0} = \sum_{\vec{k}} c^\dagger_{\vec{k}} h_{0,\vec{k}} c^\pdagger_{\vec{k}}$ with the $4\times 4$ Bloch-Hamiltonian in sublattice ($\tau_j$) and spin space ($s_j$),
\begin{equation}
\label{eq:lattice}
    h_{0,\vec{k}} = (\tau_0 \epsilon_{\vec{k}} + \tau_x t_{x,\vec{k}} + \tau_z t_{z,\vec{k}} ) s_0 + ( g_{x,\vec{k}} s_x + g_{y,\vec{k}} s_y )\tau_x.
\end{equation}
Here, $\epsilon_{\vec{k}} = -2 t_2 [\cos k_x + \cos k_y] -\mu$ is the sublattice-independent part of the dispersion, $t_{z,\vec{k}} = -2 \delta t_2 [\cos k_x - \cos k_y]$ the sublattice antisymmetric part, and $t_{x,\vec{k}} = 4 t_1 \cos \frac{k_x}{2} \cos \frac{k_y}{2}$ couples the two sublattices. The two contributions $g_{x,\vec{k}}  = 2\alpha_R \cos \frac{k_x}{2} \sin \frac{k_y}{2}$ and $g_{y,\vec{k}}  = -2\alpha_R \cos \frac{k_y}{2} \sin \frac{k_x}{2}$ lock the spin and spatial degrees of freedom.
The term describing long-range magnetism is given by $H_{\text{m}} = J\sum_{\vec{R}} c^\dagger_{\vec{R}} \vec{s}\tau_z c^\pdagger_{\vec{R}} \cdot \hat{\vec{n}}(\vec{R})$, where the unit vector $\hat{\vec{n}}(\vec{R})$ defines the local orientation of the staggered magnetization. As expected, $h_{0,\vec{k}}$ commutes with $\tau_x$ for $\delta t_2 = 0$, which corresponds to the presence of a translational symmetry relating the red and blue sublattices in \figref{fig:1}(a).

\begin{figure}
    \centering
    \includegraphics[width=\linewidth]{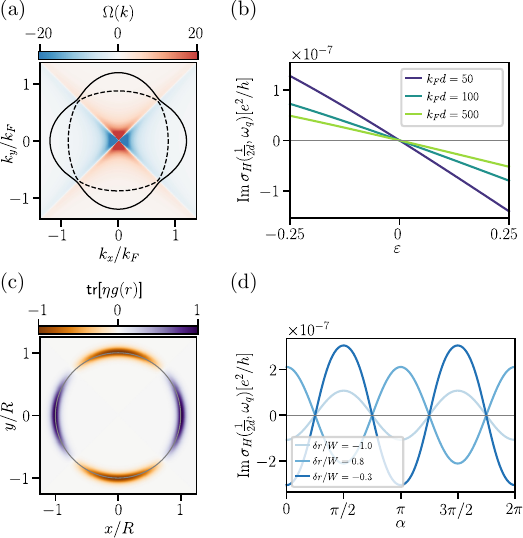}
    \caption{(a) Berry curvature for the AM. The solid and dashed lines show the corresponding Fermi surfaces. (b) Imaginary part of the Hall conductance as a function of axial strain (see text). Here, we used the Poisson ration $\nu = 0.25$. (c) Spatial profile of the (normalized) effective potential due to a domain wall. (d) Imaginary part of the Hall conductance at different positions (angles) of the domain wall for a qubit at $k_F d=100$. We used a $k_F W = 10$ and $R/W = 25$. In all plots, we considered $\mu=0.3$, $\delta t_2 = 0.3$, $\alpha_R =0.02$ and $\omega_q=10^{-4}$. Note that the relevant contribution to the decoherence difference $\delta \Gamma$ comes from momenta $\vec{q}$ with $|\vec{q}| = 1/2d$.}
    \label{fig:hall}
\end{figure}

Let us first focus on spatially uniform magnetism, $\hat{\vec{n}}(\vec{R}) = \hat{\vec{e}}_z$. If the strength $J$ of the altermagnetic order parameter is large enough, there is a gap between the two lower and the two upper bands. For chemical potentials within the energetic range of, say, the lower pair of bands, we obtain the two Fermi surfaces shown in \figref{fig:hall}(a), which are split as a result of the altermagnetic order parameter and spin-orbit coupling. To understand the origin better, we can project the Bloch Hamiltonian $h_{0,\vec{k}} + J \tau_z s_z$ onto the lower eigenspace of $J \tau_z s_z$, where spin and sublattice are anti-correlated, $\tau_z s_z = -1$. Denoting the effective Pauli matrices describing this spin-sublattice locked $2 \times 2$ subspace by $\sigma_j$, the resulting effective two-band model reads as
\begin{equation}
    h_{\text{eff},\vec{k}} = \sigma_0 \frac{\vec{k}^2}{2m} - \delta t_2 \sigma_z (k_x^2-k_y^2) + \alpha_R (k_y\sigma_x - k_x\sigma_y), \label{EffectiveHamiltonian}
\end{equation}
where we have already performed a gradient expansion---only keeping terms up to leading order in momentum in each term---and set $(2m)^{-1} = t_2$. The continuum approximation will be essential for our discussion of spatially varying altermagnetic textures in \secref{DomainWalls} below. Equation~(\ref{EffectiveHamiltonian}) makes the altermagnetic spin splitting manifest in the Hamiltonian and the fact that a finite hopping imbalance $\delta t_2 \neq 0$ is essential for this effect; it vanishes for $\delta t_2 = 0$ since the two sublattices are related by translation in this limit and the magnetic order reduces back to antiferromagnetism.

Similar to many of the features in the noise of \secref{Insulating} related to magnons, we also expect that there will be quantitative differences in the noise spectrum when comparing AM and AFM on the level of \equref{EffectiveHamiltonian}. We, however, focus here on more striking symmetry signatures that, if observed, would allow us to uniquely distinguish between AMs and AFMs.

Although it is not directly the Berry curvature $\Omega(\vec{k})$ that will play the central role, we can still use it to illustrate the key symmetry argument, as $\Omega(\vec{k})$ provides a way of quantifying the degree (and sign) of broken time-reversal symmetry in the orbital channel at momentum point $\vec{k}$. We obtain $\Omega(\vec{k}) = 0$ for the AFM limit $\delta t_2 = 0$; meanwhile, the Berry curvature is non-zero for the AM as is also shown in \figref{fig:hall}(a). It still averages to zero, $\sum_{\vec{k}} \Omega(\vec{k}) = 0$, as a result of $C_{4z}$ rotational symmetry combined with time reversal $\Theta = i \sigma_y \mathcal{K}$ ($\mathcal{K}$ is complex conjugation), making it not directly visible in isotropic local probes like NV noise spectroscopy. However, when perturbing the system in a way that breaks $C_{4z}$, the crucial difference between AM and AFM will become visible, as we discuss next.

\subsection{Strain}\label{Strain}
The first type of such rotational-symmetry-breaking perturbation we will consider is strain, which may either be applied uniformly and intentionally in experiment (see, e.g., \cite{StrainAltermagnet,PhysRevLett.125.147001} in the context of AM candidate materials) or be naturally present locally in different regions of the sample. Focusing on uniaxial strain, the associated ($2 \times 2$) strain tensor $\mathcal{E} = R^T_{\varphi} \text{diag}(-\varepsilon,\nu\varepsilon)R_{\varphi}$, with $R_\varphi$ denoting the orthogonal matrix rotating two-component vectors by an angle $\varphi$, depends on two independent parameters---the strength $\varepsilon$ and direction $\varphi$; the Poisson ratio $\nu$, in turn, is a material-dependent constant. 

On the level of the model in \equref{EffectiveHamiltonian}, we incorporate the effect of strain by bond stretching. We picture the original lattice model, \equref{eq:lattice}, being distorted along the $x$ axis ($\varphi=0$ above) such as the lattice constants become direction dependent. Hence the couplings $t_2, \delta t_2, \alpha_R$ become anisotropic, too; for instance, $t_2 \vec{k}^2 \rightarrow t_2 [(1+\varepsilon)k_x^2 + (1-\nu\varepsilon)k_y^2]$. The remaining couplings transform in the same fashion. This clearly breaks the $C_{4z}\Theta$ and the reflection symmetry $\sigma_d$ along the diagonal. As such, there is no symmetry anymore forcing the angular average of $\Omega(\vec{k})$ to vanish.

To probe this symmetry reduction, which we just illustrated using $\Omega(\vec{k})$ as an example, in the noise spectrum, we follow \refcite{9h4l-21mt} and compare $T_{1}(\hat{\vec{n}}_{\text{q}}=\hat{\vec{z}})$ with $T_{1}(-\hat{\vec{z}})$. Note that
\begin{equation}
\begin{split}
    \frac{1}{T_1(\hat{\vec{n}}_{\text{q}})} &\propto \widetilde{\mathcal{N}}_{xx}(\omega_{\text{q}}) + \widetilde{\mathcal{N}}_{yy}(\omega_{\text{q}}) \\ &
    + \,\text{Im}[\widetilde{\mathcal{N}}_{xy}(\omega_{\text{q}}) - \widetilde{\mathcal{N}}_{yx}(\omega_{\text{q}})  ],
    \end{split}
\end{equation}
where $\widetilde{\mathcal{N}} = R_{\hat{\vec{n}}_{\text{q}}}\mathcal{N} R^T_{\hat{\vec{n}}_{\text{q}}}$ is the rotated noise tensor, with $R_{\hat{\vec{n}}_{\text{q}}}$ rotating $\hat{\vec{z}}$ onto the qubit quantization axis $\hat{\vec{n}}_{\text{q}}$. As such, the difference $\delta\Gamma = T^{-1}_{1}(\hat{\vec{z}}) - T^{-1}_{1}(-\hat{\vec{z}})$ is proportional to $\text{Im}[\mathcal{N}_{xy}(\omega_{\text{q}}) - \mathcal{N}_{yx}(\omega_{\text{q}})]$. As time-reversal $\Theta$ acts as complex conjugation on $\mathcal{N}$ (see \appref{SymmetriesOnN}), $\delta\Gamma \neq 0$ implies that $\Theta$ is broken as well as $C_{4z}\Theta$ and $\sigma_d$. Adding strain to \equref{EffectiveHamiltonian}, all of these symmetries are broken. However, in the AFM limit, we still obtain $\Theta$ combined with a lattice translation as a symmetry, which implies $\delta\Gamma \rightarrow 0$ (at least for $d \gg a$).

To demonstrate this, we explicitly compute $\delta\Gamma$ with noise coming from the current fluctuations coupling to the qubit through \equref{BiotSavart}. As expected from being proportional to $\text{Im}[{\mathcal{N}}_{xy}(\omega_{\text{q}})-{\mathcal{N}}_{yx}(\omega_{\text{q}})]$ it is related to the Hall conductivity $\sigma_{\text{H}}$ of the system; as shown explicitly in \refcite{9h4l-21mt}, it holds
\begin{equation}
    \delta\Gamma \approx \frac{\mu_0^2 \omega_{\text{q}} \coth \beta \omega/2}{16 \pi^2 } \int \diff q \,q  e^{- 2qd} \int \diff \varphi \,\text{Im} \left[\sigma_{\text{H}}(\vec{q},\omega_{\text{q}})\right], \label{DeltaGamma}
\end{equation}
where $\vec{q} = q (\cos \varphi,\sin \varphi)^T = q \hat{\vec{e}}_\varphi$. We here neglected the momentum dependence of the screening, as this does not crucially alter our results, and focused on the limit ($d \gg a$) where no sublattice resolution needs to be taken into account in the Fourier transform. For an effectively non-interacting model, like in \equref{EffectiveHamiltonian}, we can compute the Hall conductivity as $\sigma_H = \frac{1}{2}(\sigma_{xy} - \sigma_{yx})$ and
\begin{widetext}
\begin{equation}
\sigma_{ij}(\mathbf{q},\omega_q)
= \frac{i e^2}{\hbar(\omega+i\eta)}
  \frac{1}{(2\pi)^2}
  \int d^2k\,
  \sum_{n,m}
  \frac{f(E_n^{\mathbf{k}}) - f(E_m^{\mathbf{k}+\mathbf{q}})}
       {E_n^{\mathbf{k}} - E_m^{\mathbf{k}+\mathbf{q}}
        + \hbar\omega_q + i\hbar\eta}\;
  \bigl[v_i(\mathbf{k},\mathbf{q})\bigr]_{nm}\,
  \bigl[v_j(\mathbf{k},\mathbf{q})\bigr]_{nm}^*, \label{HallConductivity}
\end{equation}
\end{widetext}
where $[v_\alpha(\mathbf{k},\mathbf{q})]_{nm}
= \langle n\mathbf{k}|\, V_\alpha(\mathbf{k},\mathbf{q})\, |m,\mathbf{k}{+}\mathbf{q}\rangle$ with the velocity operator,
$
    V_\alpha(\mathbf{k}, \mathbf{q}) = (
  \frac{\partial h_{\text{eff},\vec{k}}}{\partial k_\alpha}\big|_{\mathbf{k}}
  + \frac{\partial h_{\text{eff},\vec{k}}}{\partial k_\alpha}\big|_{\mathbf{k}+\mathbf{q}}
)/2
$, and $f(E)$ is the Fermi-Dirac distribution. In our numerical evaluation of the expression in \equref{HallConductivity}, we will use a small broadening $\eta$.

As already mentioned above, the factor $q  e^{-2 qd}$ in \equref{DeltaGamma} is peaked at $q = \frac{1}{2d}$, so that $\delta\Gamma$ is primarily determined by $\braket{\text{Im} \left[\sigma_{\text{H}}(\frac{1}{2d},\omega_{\text{q}})\right]} := \int \frac{\diff \varphi}{2\pi} \text{Im} \left[\sigma_{\text{H}}(\frac{\hat{\vec{e}}_\varphi}{2d},\omega_{\text{q}})\right]$. 
As before in \secref{Insulating}, this alludes to small momenta with respect to the Brillouin zone for realistic distances $d$. However, we are now probing the behavior of electrons, and $\vec{q}$ is the momentum transfer relative to their Fermi surface, which is generically at sizable momenta and, thus, sensitive to the altermagnetic symmetry reduction. Indeed, by computing $\sigma_{\text{H}}$ from the model in \equref{EffectiveHamiltonian}, with the result shown in \figref{fig:hall}(b), we can clearly see how local strain $\varepsilon$ can induce such an asymmetry, already in the limit $d\gg a$. As is clear by symmetry (translations combined with $\Theta$), this effect is not present in AFMs, as we have also checked explicitly within our model. We note that the values for $k_F d$ are chosen in our numerics so $d$ corresponds to values in the 10-100 nm range, for $k_F \approx \sqrt{\mu} = 5\, \text{nm}^{-1}$.

Note that the symmetry considerations that allow us to connect time-reversal symmetry breaking to the imaginary part of cross terms in the noise tensor are general and do not depend on the type of system studied. In particular, they also apply to the magnons studied in \secref{Insulating}. However, we note that $\delta \Gamma$ quickly vanishes as a function of distance $d/a$ within our Holstein-Primakoff approach. Only for $d \lesssim a$, where the difference between the lattice and continuum Fourier-transform of the dipole tensor matters, $\delta \Gamma$ can be probed. But this would require bringing the NV qubit very close to the sample surface. We detail these aspects in \appref{app:relaxation}.

\subsection{Domain walls}\label{DomainWalls}
Another class of perturbations that reduce the symmetries locally and will allow us to distinguish between AMs and AFMs are domain walls in the order parameter $\hat{\vec{n}}(\vec{r})$. They are ubiquitous and important in magnets (including AMs \cite{dp7v-qszq,NanoscaleImaging,DomainWallsSinova, PhysRevLett.133.196701,PhysRevLett.134.176401,PhysRevB.111.064422,2026arXiv260220236S}), and their impact on the electronic properties of AMs was recently studied in \cite{2026arXiv260220236S} in a related two-sublattice model. 

To leading (zeroth) order in $1/J$ and focusing on coplanar textures $\hat{\vec{n}}(\vec{r})$, the impact of spatially varying magnetic order parameters is described by the additional terms,
\begin{equation}
    \Delta h_{\text{eff}} = \frac{1}{2m} \text{tr}[g(\vec{r})]\sigma_0 - \delta t_2 \text{tr}[\eta g(\vec{r})] \sigma_z \label{AdditionalContributionToHamiltonian}
\end{equation}
to the effective two-band Hamiltonian in \equref{EffectiveHamiltonian}.
Here $g_{ij}(\vec{r}) = \partial_i \hat{\vec{n}} \cdot \partial_j \hat{\vec{n}}/4$ encodes the spatial variation of the texture and can be thought of as the quantum metric of the associated spin-$1/2$, two-level system. Furthermore, $\eta=\text{diag}(1,-1)$ is the metric tensor of the $d$-wave AM considered, since $k_x^2 -k_y^2 = \vec{k}^T \eta \vec{k}$.
 
While the first of the two terms in \equref{AdditionalContributionToHamiltonian} simply renormalizes the chemical potential and, hence, does not affect the symmetries of the system, the second one can be thought of as an emergent local Zeeman field. For a circular domain wall, it inherits the $d$-wave nature of the AM, as can be seen in \figref{fig:hall}(c). Since this effect vanishes in the AFM limit, $\delta t_z \rightarrow 0$, it is a hallmark of the AM, as pointed out in \refcite{2026arXiv260220236S}. Here, we show that this not only leads to signatures in the local magnetization but also in the noise.

First, notice that the additional term $\propto \sigma_z$ breaks both $C_{4z}\Theta$ and $\sigma_d$ and should thus induce finite $\delta\Gamma$. This is indeed the case, as can be seen in our results presented in \figref{fig:hall}(d), where we show the angular dependence at a fixed distance from the domain wall. We can further see that the sign of $\delta\Gamma$ follows the $d$-wave nature of $\text{tr}[\eta g(\vec{r})]$ coming from the $k_x^2-k_y^2$ dependence of the projected altermagnetic order parameter. Since a $g$-wave AM will also give rise to a $g$-wave pattern in $\text{tr}[\eta g(\vec{r})]$ along a domain wall \cite{2026arXiv260220236S} and due to the locality of \equref{AdditionalContributionToHamiltonian}, $\delta\Gamma$ will exhibit a $g$-wave behavior too. This shows that the noise spectrum along a domain wall not only differs qualitatively between AMs and AFMs, but also provides access to the dominant orbital nature of the AM.

Apart from $\Delta h_{\text{eff}}$ in \equref{AdditionalContributionToHamiltonian}, there is a second modification in the Hamiltonian that has already been taken into account in \figref{fig:hall}(d). To understand its origin and form, let us first recall how $\Delta h_{\text{eff}}$ arises.
We obtain $\Delta h_{\text{eff}}$ by rotating the effective Hamiltonian in \equref{EffectiveHamiltonian} into the co-rotating reference frame \cite{Volovik1987,2026arXiv260214950M,2026arXiv260220236S}, i.e., $U^\dagger(\vec{r}) \vec{n}(\vec{r}) \cdot \vec{\sigma} U(\vec{r}) = \sigma_z$. Since we start with a Hamiltonian with spin-orbit coupling, the orientations in spin space and real space are locked. Consequently, we distinguish between Bloch-like and N\'eel-like domain walls. For the former, a natural choice would be $U=\exp(i\frac{\phi(r)}{2} (\sigma_x \cos\alpha + \sigma_y \sin\alpha))$, where $\phi(r)$ interpolates between $0$ and $\pi$ and $\alpha$ is the (direction) polar angle. In this case, $\vec{n}(\vec{r})$ rotates perpendicular to the domain wall. For a N\'eel-like domain wall, the order rotates parallel to the domain wall, corresponding to $U=\exp(i\frac{\phi(r)}{2} (-\sigma_x \sin\alpha + \sigma_y \cos\alpha))$. We here neglect the non-coplanarity of the resulting texture (no derivative with respect to $\alpha$ entering $g$), which makes sense in the limit of small curvature of the domain wall (in our case, $R\rightarrow \infty$). In this limit, one can think of $\alpha$ also simply as the orientation of a straight domain wall with respect to the lattice/altermagnetic order parameter. An important consequence is that $U^\dagger \vec{\nabla} U$ is entirely off-diagonal in spin space and, thus, vanishes when projected to the low-energy subspace with fixed $\tau_z s_z$. As such, the emergent gauge fields \cite{Volovik1987,2026arXiv260214950M,2026arXiv260220236S} (and not only the associated orbital magnetic field) vanish for these specific choices of $U$.

However, in addition to the emergent Zeeman field in \equref{AdditionalContributionToHamiltonian}, the rotation $U(\vec{r})$ also transforms the spin-orbit coupling in \equref{EffectiveHamiltonian} according to $\sigma_{x,y} \rightarrow \widetilde{\sigma}_j(\vec{r})$ where
\begin{equation}
    \widetilde{\sigma}_j(\vec{r}) = \sum_{j'=x,y} \sigma_{j'}\text{tr}[U^\dagger(\vec{r}) \sigma_j U(\vec{r}) \sigma_{j'}]/2,    
\end{equation}
see \appref{TransformationSOC}. The consequences of the modifications to the spin-orbit coupling term can be seen especially right at the domain wall ($\delta r =0$). There, we find the largest effective potential $\Delta h_{\text{eff}}$. However, we also have $\phi(R) = \pi/2$ and the $\hat{\vec{n}}$ is in the plane of the system. This implies that the spin-orbit coupling term becomes $\alpha (\hat{\vec{n}}_{\perp}\cdot \vec{k} ) (\vec{\sigma} \cdot \hat{\vec{n}})$, where $\hat{\vec{n}}_{\perp} = \hat{\vec{z}} \times \hat{\vec{n}}$. Consequently, the Bloch vector is confined to a plane (spanned by $\hat{\vec{n}}$ and $\hat{\vec{z}}$) across all momenta, which gives a vanishing Berry curvature. Therefore, right at the wall, when the effective potential is the largest, we have $\sigma_H =0$. This results in a sign change of the $\delta \Gamma$ through the domain wall (at fixed angle $\alpha$), i.e., when increasing $r$ from $r<R$ to $r>R$, which is also visible in \figref{fig:hall}(d). This shows that the noise through $\delta \Gamma$ is expected to display a qualitatively different signature than $V_z$, which is a consequence of the fact that the former is determined by both the emergent Zeeman field $V_z$ and the (renormalized) spin-orbit coupling. 

\subsection{Symmetry considerations}
\label{SymmetryConsiderations}

Hitherto, we mainly discussed the difference in the relaxation rate of anti-aligned qubits with $\vec{n}_{\text{q}} = \pm \hat{\vec{z}}$. Similarly to what we presented in \secref{MagneticNoise}, and more specifically in \figref{fig:angular}, one can investigate the more general dependence of the relaxation and decoherence rates on the qubit orientation $\hat{\vec{n}}_{\text{q}}$ for itinerant electrons as well. 
As we have seen with $\delta \Gamma$ as a concrete example above, the finite size of Fermi surfaces makes the noise related to the itinerant electrons generically susceptible to any symmetry reduction in the system. We therefore discuss additional features of these quantities based on general symmetry arguments.

The two equations that define the symmetry constraints for the noise tensor are
\begin{equation}
\begin{split}
    &\mathcal{N}(\Omega) = R^T(g)\mathcal{N}(\Omega) R(g), \\
    &\mathcal{N}(\Omega) = R^T(g)\mathcal{N}^T(\Omega) R(g).
\end{split}
\end{equation}
In the first equation, $R(g)$ is the appropriate pseudo-vector representation of the point-group symmetry $g$. In the second equation, the consequence of a magnetic symmetry is considered, i.e., $g \Theta$ with time-reversal $\Theta$. A detailed discussion regarding these equations can be found in \appref{SymmetriesOnN}.

Now, we can take into account the appropriate symmetries for an AFM and an AM. For concreteness, let us assume the magnetic order is along the $\hat{\vec{y}}$ direction, that spin-orbit coupling is present, and that a magnetic field $B$ is applied along the $\hat{\vec{z}}$ direction. 
Then a square-lattice AFM has the magnetic mirror symmetry $\sigma_y \Theta$ and $C_{2z}$, while an AM only has $\sigma_y \Theta$. We emphasize that all symmetries here are understood as spinful transformations, where not only the lattice coordinates but also the spins are transformed accordingly. In the first case, this implies that
\begin{equation}
    \mathcal{N}_{\text{AFM}} = \left(
\begin{array}{ccc}
 \mathcal{N}_{xx} & \mathcal{N}^B_{xy} & 0 \\
  -\mathcal{N}_{xy}^B & \mathcal{N}_{yy} & 0 \\
 0 & 0 & \mathcal{N}_{zz} \\
\end{array} 
\right) \label{NoiseTensorAFMConstr}
\end{equation}
where the off-diagonal term is purely imaginary and only present when $B\neq 0$, as indicated by the superscript. More precisely, we use the notation that all components with (without) superscript $B$ are odd (even) in $B$. This behavior follows from the presence of the symmetry given by translation combined with $\Theta$ and $B \rightarrow -B$ in the AFM. Importantly, this symmetry (just like $C_{2z}$) is broken in the AM. More terms are possible and one finds
\begin{equation}
    \mathcal{N}_{\text{AM}} = \left(
\begin{array}{ccc}
 \mathcal{N}_{xx} & \mathcal{N}_{xy}^B & \mathcal{N}_{xz} \\
  -\mathcal{N}_{xy}^B & \mathcal{N}_{yy} & \mathcal{N}^B_{yz} \\
 -\mathcal{N}_{xz} & \mathcal{N}^B_{yz} & \mathcal{N}_{zz} \\
\end{array}
\right). \label{NoiseTensorAMConstr}
\end{equation}
Similarly to the AFM, $\mathcal{N}^B_{xy}$ and $\mathcal{N}_{xz}$ are purely imaginary. The remaining off-diagonal term, $\mathcal{N}^B_{yz}$, is real.
Note that, in principle, these statements are also true for the calculations in \secref{Insulating}; however, the dipolar magnetic field and the magnon spectra within our leading-order Holstein-Primakoff approach impose further constraints and affect the angular dependence of the noise.

Using the form of the noise tensor in \equsref{NoiseTensorAFMConstr}{NoiseTensorAMConstr}, one can readily derive the general form of $1/T_{1,2}(\vec{n}_{\text{q}})$, see \appref{SymmetriesOnN}. First, we find that $\delta \Gamma = 0$ if $B=0$ for both AFM and AM; just like in our analysis above, additional perturbations are needed to contrast the two types of orders (we also see that applying $B$ would not help since then $\delta \Gamma \neq 0$ for both).
However, we notice the presence of the anti-symmetric components $\mathcal{N}_{xz}$ in the AM, which are absent for the AFM. This term can be probed in the noise, e.g., from the rotated asymmetry term $\widetilde{\delta \Gamma} = T^{-1}_{1}(\hat{\vec{y}}) - T^{-1}_{1}(-\hat{\vec{y}})$: one finds $\widetilde{\delta \Gamma} \propto \mathcal{N}_{xz} \neq 0$ for the AM (even at $B=0$) while  $\widetilde{\delta \Gamma} = 0$ for the AFM. But we emphasize that $\mathcal{N}_{xz}$ also enters for other generic directions of $\vec{n}_{\text{q}}$. Most importantly, this shows that the noise also allows access to unique symmetry signatures of AM even without additional perturbations (like the strain or domain walls discussed above). 

Finally, we point out that there are also unique signatures of AMs in $T_2$. As follows from \equsref{NoiseTensorAFMConstr}{NoiseTensorAMConstr}, $1/T_{2}(\vec{n}_{\text{q}})$ has a $B$-linear contribution (for generic $\vec{n}_{\text{q}}$) only in the AM, which is related to the $\mathcal{N}^B_{yz}$ term and is thus absent for the AFM. This is also the only term (and thus only non-vanishing in the AM) leading to differences in $1/T_{2}(\vec{n}_{\text{q}})$ for $\phi$ and $\phi+\pi$ [cf.~parameterization in \secref{MagneticNoise}].

\section{Conclusion}\label{Conclusion}
In this work, we explored the potential of noise magnetometry, e.g., as implemented via NV centers, to probe and characterize altermagnets. We have calculated and compared the magnetic noise spectrum of an antiferromagnet and altermagnet, realized in the two-sublattice model illustrated in \figref{fig:1}(a). We considered both the fluctuations in the magnetic order parameter (magnons) as well as the charge fluctuations of itinerant electrons in the bands that are reconstructed by the magnet.

We first focused on the fluctuations related to magnons, which are the central contribution for insulating systems. We employed a Heisenberg model and studied the regime with gapless magnons, i.e., when the noise is dominated by single-magnon processes. We identified quantitative differences between altermagnets and antiferromagnets in the noise spectrum. While in principle key features of the magnon bandstructure---the splitting of the bands in the altermagnetic case---can also be seen in the noise spectrum, it requires access to large energy ranges and the qubit being very close to the sample. We note that this regime might be accessible in moir\'e systems \cite{PhysRevLett.133.206702,2026arXiv260219734R}, where the effective energy scales and lattice constants are reduced. We discussed the relevance of taking into account lattice effects in the Fourier transform of the dipolar interaction kernel in this regime.

In the second part, we studied the additional contribution of charge fluctuations relevant to itinerant altermagnets, which we captured starting from a tight-binding model. This is a particularly important contribution since the non-trivial impact of the magnetic order parameter on the itinerant charge carriers is the central feature of altermagnets and the Fermi surface is typically at sizable momenta, making the noise spectra generically susceptible to altermagnetic symmetries. To illustrate this, we computed $[1/T_1 (\hat{\vec{z}}) - 1/T_1 (-\hat{\vec{z}})] \propto \sigma_H(\omega_q)$ in the presence of distortions that lift the rotational symmetry $C_{4z}$ and the mirror symmetry $\sigma_d$. We studied uniaxial strain and domain walls as such symmetry-breaking distortions; this allows a finite $\sigma_H$ only for the altermagnet, while antiferromagnets show no finite response, therefore, giving us a qualitative difference between the two phases. What is more, the angular dependence of the noise around domain walls further contains information about the dominant orbital character of the altermagnet. We showed by a symmetry analysis that there are also contributions in the noise spectrum that distinguish altermagnets from antiferromagnets without additional perturbations, such as $T^{-1}_{1}(\hat{\vec{y}}) - T^{-1}_{1}(-\hat{\vec{y}})$ (assuming in-plane magnetic order).

Overall, our work shows that noise magnetometry can provide crucial insights into altermagnetism, in particular in the itinerant case. It further leads to many natural follow-up questions, such as the impact not only of static strain but also of fluctuations in it \cite{PhysRevB.108.144418,1vqq-9kzm}. It would also be interesting to generalize our symmetry analysis and model calculations to altermagnetic and unconventional magnets on other Bravais lattices, and take into account the emergent spin-orbit coupling in inhomogeneous altermagnets \cite{2026arXiv260220236S,2026arXiv260214950M}.

\begin{acknowledgments}

We thank R.~Peng, S.~Banerjee, C.~Schrade, U.~Seifert, J.~A.~Sobral, and J.~Wrachtrup for insightful discussions.
We further acknowledge funding by the European Union (ERC-2021-STG, Project 101040651— SuperCorr). Views and opinions expressed are however those of the authors only and do not necessarily reflect those of the European Union or the European Research Council Executive Agency. Neither the European Union nor the granting authority can be held responsible for them.
\end{acknowledgments}

\bibliography{draft_Refs}

\onecolumngrid

\begin{appendix}

\section{Holstein-Primakoff transformation in the local basis}
\label{app:h-p}
As mentioned in the main text, we parametrize $\vec{S}_i = \vec{m}_i S^\parallel + \vec{S}^\perp_i$, with $\vec{m}_i \cdot \vec{S}^\perp_i = 0$, and rewrite the Heisenberg Hamiltonian as

\begin{equation}
    H = J_1 \sum_{\langle i,j\rangle} \left[(\vec{m}_i\cdot \vec{m}_j) S_i^\parallel S_j^\parallel + \vec{S}_i^\perp \cdot \vec{S}_j^\perp \right]+ J_2 \sum_{\langle\langle i,j\rangle\rangle} \left[(\vec{m}_i\cdot \vec{m}_j) S_i^\parallel S_j^\parallel + \vec{S}_i^\perp \cdot \vec{S}_j^\perp \right] - \sum_i \vec{h}\cdot \vec{m}_i \,\,S^\parallel  + \mathcal{O}(1/S^0)\,\,\,.
\end{equation}
Within our $1/S$ expansion, we can neglect terms with $\vec{B}\cdot \vec{S}_i^\perp $ and $\vec{S}_i^\perp \cdot S_j^\parallel$. Now, we use the Holstein-Primoakoff transformation in the local basis $(\vec{e}_{1,i},\vec{e}_{2,i}, \vec{m}_i)$ (also defined in the main text and see \cite{PhysRevLett.91.017205}) and obtain the ``classical'' contribution to the Hamiltonian
\begin{equation}
    H_{\text{cl}} = J_1 S^2 \sum_{\langle i,j \rangle} \vec{m}_i\cdot \vec{m}_j + J_2 S^2 \sum_{\langle \langle i,j \rangle\rangle} \vec{m}_i\cdot \vec{m}_j - S \sum_i \vec{h}\cdot \vec{m}_i = \mathbbm{1} E_{\text{cl}}[\{\vec{m}_i\}]
\end{equation}
and the $\mathcal{O}(S)$ correction, which is quadratic in the bosonic operators and reads as
\begin{equation}
    \begin{split}
        \mathcal{H}/S &=J_1 \sum_{\langle i,j \rangle} \left[- \vec{m}_i\cdot \vec{m}_j (n_i + n_j) +2 (\vec{e}_i^+ \cdot \vec{e}_j^+) b_i^\pdagger b_j^\pdagger +2 (\vec{e}_i^+ \cdot \vec{e}_j^-) b_i^\pdagger b_j^\dagger +2 (\vec{e}_i^- \cdot \vec{e}_j^+) b_i^\dagger b_j^\pdagger + 2 (\vec{e}_i^- \cdot \vec{e}_j^-) b_i^\dagger b_j^\dagger\right]\\
        &+J_2 \sum_{\langle \langle i,j \rangle \rangle}\left[- \vec{m}_i\cdot \vec{m}_j (n_i + n_j) +2 (\vec{e}_i^+ \cdot \vec{e}_j^+) b_i^\pdagger b_j^\pdagger +2 (\vec{e}_i^+ \cdot \vec{e}_j^-) b_i^\pdagger b_j^\dagger +2 (\vec{e}_i^- \cdot \vec{e}_j^+) b_i^\dagger b_j^\pdagger + 2 (\vec{e}_i^- \cdot \vec{e}_j^-) b_i^\dagger b_j^\dagger \right]\\
        &+ \sum_i \vec{h}\cdot \vec{m}_i \,\,n_i,
    \end{split}
\end{equation}
where we introduced $n_i = b_i^\dagger b^\pdagger_i$.

For our choice of parameters ($\vec{h}=h \hat{z}$; $J_1 >0$ dominating the energetic preference of $J_2$), $E_{\text{cl}}[\{\vec{m}_i\}]$ is minimized by a magnetization of the form $\vec{m}_i = \varrho_i \cos{\theta} \,\, \hat{x} + \sin{\theta} \hat{z}$ with $\varrho= \pm1$ depending on the sublattice. The classical part becomes
\begin{equation}
    H_{\text{cl}} = N S\,\,(- h \sin{\theta} - J_1 S \cos{2 \theta} + J_2 S).
\end{equation}
The minimum is reached for 
\begin{equation}
\theta_{m} \;=\;
\begin{cases}
\arcsin\!\left(\dfrac{h}{8 J_1 S}\right), & |h| \leq 8 J_1 S, \\[10pt]
\dfrac{\pi}{2}, & |h| > 8 J_1 S .
\end{cases}
\end{equation}

We can now write down the explicit local basis
\begin{equation}
    \begin{split}
        \vec{e}_{1,i} &= \varrho_i \sin{\theta_m} \,\,\hat{x} - \cos{\theta_m} \,\,\hat{z};\\
        \vec{e}_{2,i} &= \varrho_i\,\, \hat{y};\\
    \vec{m}_i &= \varrho_i \cos{\theta_m} \,\, \hat{x} + \sin{\theta_m} \hat{z}.
    \end{split}
\end{equation}
The dot products in $\mathcal{H}$ are

\begin{equation}
    \begin{split}
        \vec{e}^\pm_i \cdot \vec{e}^\pm_j &= \begin{cases}
            0, \quad \quad \quad \quad\quad\,\,\, \text{if $i,j$ are in the same sublattice,}\\
            \frac{\cos^2{\theta_m}}{2} \quad \quad \text{if $i,j$ are in different sublattice,}
        \end{cases}\\
        \\
        \\
         \vec{e}^\pm_i \cdot \vec{e}^\mp _j &= \begin{cases}
            \frac{1}{2}, \quad \quad \quad \quad\quad\,\,\, \text{if $i,j$ are in the same sublattice,}\\
            -\sin^2{\theta_m} \quad \quad \text{if $i,j$ are in different sublattice,}
        \end{cases}\\
        \\
        \\
        \vec{m}_i\cdot \vec{m}_j &= \begin{cases}
            1, \quad \quad \quad \quad\quad\,\,\, \text{if $i,j$ are in the same sublattice,}\\
            -\cos{2 \theta_m} \quad \quad \text{if $i,j$ are in different sublattice.}
        \end{cases}\\ 
    \end{split}
\end{equation}
Hence the Hamiltonian is

\begin{equation}
    \begin{split}
        \mathcal{H}/S &=J_1 \sum_{\langle i,j \rangle} \left[ \cos{2 \theta_m}\,\, (n_i + n_j)  - \sin^2{\theta_m}\,\, (b_i^\pdagger b_j^\dagger +b_i^\dagger b_j^\pdagger) + \cos^2{\theta_m}\,\,( b_i^\dagger b_j^\dagger+ b_i^\pdagger b_j^\pdagger) \right]\\
        &+J_2 \sum_{\langle \langle i,j \rangle \rangle} \left[-  (n_i + n_j)  +\,\ b_i^\pdagger b_j^\dagger +\,\ b_i^\dagger b_j^\pdagger \right]\\
        &+ \sum_i \frac{h}{S} \sin{\theta_m} \,\,n_i.
    \end{split}
\end{equation}
Now, if we use the Nambu basis defined in the main text, we obtain  directly the matrix form in \equref{eq:h_k}.

\section{Colpa method}
\label{app:colpa}
Within the linear spin wave approximation, we want to numerically obtain the Bogoliubov coefficients and spectrum. However they are too complicated to be written in a reasonable analytical form. Hence, we need to numerically diagonalize $h_{\vec{k}}$ from \eqref{eq:h_k}, a task that is not as straightforward as it is with fermions due to the bosonic commutation relations.

A common way to proceed is through the Colpa method \cite{COLPA1978327}. The first step is to define the metric of our system $\Sigma = \text{diag}(1,1,-1,-1)$. Now we write down our Hamiltonian as $h = K^\dagger K$ (suppressing the momentum dependence in the following). This form is always possible, given that $h$ is positive definite. For our case, at $k=0$, we always have a Goldstone mode with zero energy, hence we can add a small diagonal shift, e.g., $\epsilon \sim 10^{-5}$, that makes the problem numerically stable.

The next step is to obtain $K$, which is done through a Cholesky decomposition. Then we solve the eigenproblem $K \Sigma K^\dagger$ and obtain the unitary transformation $U$ to the diagonal basis, i.e., the transformation is such that $L = U^\dagger K \Sigma K^\dagger U$ is diagonal. The eigenvalues of this problem are not the physical energies, instead $E = \Sigma L$ gives the correct energies (it transforms the negative energy bands into positive). Furthermore, the eigenvectors of the physical problem are the columns of
\begin{equation}
    T = K^{-1} U E^{1/2}.
\end{equation}
It is straightforward to check that $T \Sigma T^\dagger = \Sigma$, such that the transformation safeguards the bosonic commutation relations, while simultaneously diagonalizing the Hamiltonian, $T^\dagger h T = E$.

The transformation matrix $T$ follows the symmetries of a Bogoliubov transformation by definition and can be written as
\begin{equation}
\label{eq:bogo-transformation}
    T = \begin{pmatrix}
        \mathcal{U} & \mathcal{V}\\
        \mathcal{V} & \mathcal{U}
    \end{pmatrix},
\end{equation}
with the gauge choice where all the entries are real, which is possible because $h_k$ is real.

\section{Noise tensor expressions}
\label{app:noise}

We start with our expressions for the magnetic noise tensor in real space and the magnetic field expression for a set of dipoles
\begin{equation}
    \mathcal{N}_{\alpha \beta} (\omega) =\frac{1}{2} \int dt  e^{i \omega t } \langle \{B_\alpha(t), B_\beta(0) \} \rangle,
\end{equation}
and
\begin{equation}
\label{eq:B_app}
    \vec{B}=\frac{\mu_0 \mu_{\mathrm{B}}}{4 \pi} \sum_j\left[\frac{\vec{S}_j}{r_j^3}-\frac{3\left(\vec{S}_j \cdot \vec{r}_j\right) \vec{r}_j}{r_j^5}\right].
\end{equation}
As stated in the main text $\vec{r}_j =(x_j,y_j,d)$ corresponds to the distance between the spin in the $(x_j,y_j,0)$ position and the qubit at $(0,0,d)$. We first use the dissipation-fluctuation result $\langle\{,\} \rangle_T \rightarrow -\coth{\frac{\omega}{2T}}\,\,\theta(t)\,\, \Im \langle -i [,]\rangle$. We also note that we can rewrite the magnetic field in \equref{eq:B_app} in terms of the dipole tensor
\begin{equation}
\begin{split}
B_\alpha &= \frac{\mu_0 \mu_{\mathrm{B}}}{4 \pi}
\sum_\eta \sum_j \sum_\gamma
D^{\alpha \gamma}(\vec{r}_j + \vec{\eta})\, S^\gamma_{\eta, j}, \\
D^{\alpha \gamma}(\vec{r}_j)
&= \frac{\delta^{\alpha \gamma}}{r_j^3}
- \frac{3 r_j^\alpha r_j^\gamma}{r_j^5}\\
&= -\partial^\alpha \partial^\gamma
\left( \frac{1}{\sqrt{\rho^2 + z^2}} \right)\Bigg|_{z=d},
\end{split}
\end{equation}
where $\rho$ is the norm of the in-plane position vector $\vec{\rho}=(x,y)^T$. The index $\eta$ labels the two different sub-lattices, it also indicates the precise spin position $\vec{r}_j + \vec{\eta}$, with $\vec{\eta} = \hat{x}/2 $ or $\hat{y}/2$.

Now we can write the expectation value with a commutator
\begin{equation}
    \begin{split}
       \langle [B_\alpha(t), B_\beta(0) ] \rangle =\left(\frac{\mu_0 \mu_{\mathrm{B}}}{4 \pi}\right)^2 \sum_{\eta,\xi}\sum_{i,j} \sum_{\gamma \delta} D^{\alpha\gamma}(\vec{r}_i + \vec{\eta})  D^{\beta\delta}(\vec{r}_j +\vec{\xi})  \langle [S^\gamma_{\eta,i}(t),S^\delta_{\xi,j}(0)] \rangle,
    \end{split}
\end{equation}
and rewrite the spin operators in momentum space $S_{\eta,i}^\gamma = \frac{1}{ \sqrt{N} } \sum_{\vec{k}} e^{-i\vec{k}\cdot (\vec{\rho}_i+\vec{\eta}) } S_{\eta, \vec{k}}^\gamma$. We note that the spins ``live'' in the plane so transform $\vec{\rho}_i$ instead of $\vec{r}_i$. Now the expectation value is
\begin{equation}
    \begin{split}
       \langle [B_\alpha(t), B_\beta(0) ] \rangle =\left(\frac{\mu_0 \mu_{\mathrm{B}}}{4 \pi}\right)^2 \frac{1}{ N }\sum_{\eta,\xi}\sum_{\vec{k},\vec{k}'} \sum_{\gamma \delta} \left[ \sum_i D^{\alpha\gamma}(\vec{r}_i + \vec{\eta}) e^{-i\vec{k}\cdot (\vec{\rho}_i +\vec{\eta})}\right] \left[ \sum_j  D^{\beta\delta}(\vec{r}_j + \vec{\xi}) e^{-i\vec{k}'\cdot (\vec{\rho}_j +\vec{\xi})} \right]  \langle [S^\gamma_{\eta,\vec{k}}(t),S^\delta_{\xi,\vec{k}'}(0)] \rangle,
    \end{split}
\end{equation}
For $d \gg a$ a small-$\vec{k}$ expansion suffices (we discuss the case $d \lesssim a$ later on). Then the expressions in-between square brackets become just the 2D Fourier transform $\mathcal{D}^{\alpha\gamma}(\vec{k}) = \mathcal{F}\{D^{\alpha\gamma}(\vec r_i)\}(\vec k)$ of the dipole tensor and we can compactly write
\begin{equation}
\label{eq:ft-dipole}
    \begin{split}
       \langle [B_\alpha(t), B_\beta(0) ] \rangle =\left(\frac{\mu_0 \mu_{\mathrm{B}}}{4 \pi}\right)^2 \sum_{\eta,\xi}\sum_{\vec{k},\vec{k}'} \sum_{\gamma \delta} \mathcal{D}^{\alpha\gamma}(\vec{k}) \mathcal{D}^{\beta\delta}(\vec{k}')   \langle [S^\gamma_{\vec{k},\eta}(t),S^\delta_{\vec{k}',\xi}(0)] \rangle.
    \end{split}
\end{equation}
\begin{equation}
\mathcal{D}^{\alpha\gamma}(\vec k)
\sim
\begin{cases}
\dfrac{2\pi}{k}\,e^{-k d}\,k_\alpha k_\gamma,
& \alpha,\gamma \in \{x,y\}, \\[8pt]

\,i\,2\pi\,e^{-k d}\,k_\alpha,
& \alpha \in \{x,y\},\ \gamma = z \ \text{or}\  \alpha = z,\ \gamma \in \{x,y\}, \\[8pt]

-2\pi\,k\,e^{-k d},
& \alpha=\gamma=z .
\end{cases}
\end{equation}

We note that within this expansion, the dependence on the sub-lattices is compensated by a variable shift in the continuum Fourier transformation and therefore it does not appear explicitly in the dipole expression.

Now we take into account the translational invariance of our system (momentum conservation), $\langle [S^\gamma_{\vec{k}}(t),S^\delta_{\vec{k}'}(0)] \rangle \propto \delta_{\vec{k},-\vec{k}'}$, and go to thethermodynamic limit which allows us to replace $\frac{1}{N}\sum_{\vec{k}} \rightarrow \frac{1}{(2 \pi)^2} \int d^2\vec{k}$,
\begin{equation}
    \begin{split}
       \langle [B_\alpha(t), B_\beta(0) ] \rangle =\left(\frac{\mu_0 \mu_{\mathrm{B}}}{4 \pi}\right)^2 \sum_{\eta,\xi} \sum_{\gamma \delta} \int d^2\vec{k}  \,\,D^{\alpha\gamma}(\vec{k}) D^{\beta\delta}(-\vec{k})   \langle [S^\gamma_{\vec{k},\eta}(t),S^\delta_{-\vec{k},\xi}(0)] \rangle.
    \end{split}
\end{equation}
Hence, the noise tensor is
\begin{equation}
    \mathcal{N}_{\alpha\beta}(\omega)= \frac{1}{2} \coth{\frac{\omega}{2T}} \left(\frac{\mu_0 \mu_{\mathrm{B}}}{4 \pi}\right)^2\sum_{\gamma \delta}  \int \frac{d^2\vec{k}}{(2\pi)^2} D^{\alpha\gamma}(\vec{k}) D^{\beta\delta}(-\vec{k})  \,\,\mathcal{C}_{\gamma \delta}(\vec{k},\omega),
\end{equation}
with
\begin{equation}
    \mathcal{C}_{\gamma \delta}(\vec{k},\omega)=-\sum_{\eta,\xi}\Im (-i \int_0^\infty dt e^{i\omega t} \langle [S^\gamma_{\vec{k},\eta}(t),S^\delta_{-\vec{k},\xi}(0)] \rangle ).
\end{equation}

\subsection{Decoherence rate}
\label{appendix:subsec-decoherence}
For the decoherence rate ($1/T_2$), the important noise component is the one parallel to the qubit quantization axis. If we start with the quantization axis being along $\hat{z}$, that means the $\mathcal{N}_{zz}$ component. If we now do an aribtrary rotation $R_{\text{tot}} = R_x(\phi)R_z(\theta)$, it transforms as
\begin{equation}
\begin{split}
\mathcal{N}_{zz}\xrightarrow{R_{\text{tot}}} &\mathcal{N}_{zz}\cos^2\theta
+ \sin(2\theta)\,\sin\phi \,\,\mathcal{N}_{xz}+ \sin^2\theta \left( \mathcal{N}_{yy}\cos^2\phi + \mathcal{N}_{xx}\sin^2\phi \right).
\end{split}
\end{equation}
Here, we already omitted the components that do not contribute for the following reason. For example, the term $(\mathcal{N}_{xy} + \mathcal{N}_{yx} )\sin^{2}\theta\,\sin 2\phi$ is technically part of our expression. However, this term is
\begin{equation}
     \mathcal{N}_{xy} =\frac{1}{2} \coth{\frac{\omega}{2T}} \left(\frac{\mu_0 \mu_{\mathrm{B}}}{4 \pi}\right)^2\sum_{\gamma \delta} \int dt e^{i\omega t} \int d^2\vec{k} \,\, e^{-2kd} \frac{(k_x k_y)^2}{k^2} (\langle[S^x_{\vec{k}}(t),S^y_{-\vec{k}}(0)] \rangle + \langle[S^y_{\vec{k}}(t),S^x_{-\vec{k}}] \rangle)
\end{equation}
and upon inserting our spin representations (with the linearized Holstein-Primakoff transformation) 
\begin{equation}
    \begin{split}
            \langle[S^x_{\vec{k}}(t),S^y_{-\vec{k}}(0)] \rangle= \sum_{a,a'} i \sin \theta_m \langle\left[\frac{b_{\vec{k},a}(t) + b^\dagger_{-\vec{k},a}(t)}{2},\frac{b_{-\vec{k},a'}(0)-b^\dagger_{\vec{k},a'}(0)}{2} \right] \rangle
    \end{split}
\end{equation}
and the respective Bogoliubov coefficients, one can conclude that $\langle[S^x_{\vec{k}}(t),S^y_{-\vec{k}}(0)] \rangle= - \langle[S^y_{\vec{k}}(t),S^x_{-\vec{k}}(0)] \rangle$ and therefore $\mathcal{N}_{xy}=0$ in the continuum Fourier transform approximation. For the lattice version (relevant when $d \gg a$ does not hold), discussed later on in this appendix, $\mathcal{N}_{xy}$ is not zero due to the different dipole tensor components in its respective expressions. However $\mathcal{N}_{xy}$ is purely imaginary and, due to the relation $\mathcal{N}_{ij} = \mathcal{N}_{ji}^*$, the relevant term $(\mathcal{N}_{xy} + \mathcal{N}_{yx} )$ for $1/T_2$ is always zero. The same argument is true for the other cross component with $\mathcal{N}_{yz}$, i.e., $(\mathcal{N}_{yz} + \mathcal{N}_{zy})$ is also zero.

Now, for concreteness, let us calculate explicitly $\mathcal{N}_{zz}$. From all the combinations $D^{z\gamma}D^{z\delta} \langle[S_{\vec{k}}^\gamma(t), S_{-\vec{k}}^\delta(0)] \rangle$, only when $\gamma=\delta$ is non trivial. This happens because the spin-spin correlator is even in $k$. Hence, if the remaining terms are an odd function in $k$, that component vanishes under the $k$ integration. Therefore, we end up with

\begin{equation}
    \mathcal{N}_{zz}  =\frac{1}{2} \coth{\frac{\omega}{2T}} \left(\frac{\mu_0 \mu_{\mathrm{B}}}{4 \pi}\right)^2  \int d^2\vec{k} \,\,  e^{-2kd} k^2 \left[\mathcal{C}_{zz}(\vec{k},\omega) + \frac{\mathcal{C}_{xx}(\vec{k},\omega) + \mathcal{C}_{yy}(\vec{k},\omega)}{2}\right]
\end{equation}
Now, if we plug the spin expressions from \equref{eq:spin-hp} and the Bogoliubov coefficients from the diagonalizing transformation (\equref{eq:bogo-transformation}), we obtain
\begin{equation}
\begin{split}
   \mathcal{C}_{xx}(\vec{k},\omega) =(2 S) \, \sin^2\theta_m  \frac{1}{4}\,\left[\delta(\omega - \omega_{+,\vec{k}}) (u_{11} - u_{12} + v_{11}- v_{21})^2 \right.\\
    +\left. \delta(\omega - \omega_{-,\vec{k}}) (u_{12} - u_{22} + v_{12}- v_{22})^2  \right],
    \end{split}
\end{equation}

\begin{equation}
\begin{split}
    \mathcal{C}_{yy}(\vec{k},\omega) =(2 S) \frac{1}{4}\, \,\left[\delta(\omega - \omega_{+,\vec{k}}) (u_{11} - u_{12} - v_{11}+ v_{21})^2 \right.\\
    +\left. \delta(\omega - \omega_{-,\vec{k}}) (u_{12} - u_{22} - v_{12}+ v_{22})^2  \right],
    \end{split}
\end{equation}

\begin{equation}
\begin{split}
    \mathcal{C}_{zz}(\vec{k},\omega) =(2 S) \,\frac{1}{4} \cos^2\theta_m \,\left[\delta(\omega - \omega_{+,\vec{k}}) (u_{11} - u_{12} + v_{11}- v_{21})^2 \right.\\
    +\left. \delta(\omega - \omega_{-,\vec{k}}) (u_{12} - u_{22} + v_{12}- v_{22})^2  \right].
    \end{split}
\end{equation}
In this last step we used the time evolution of the eigenmodes ($\omega_{+,\vec{k}}$ and $\omega_{-,\vec{k}}$) of our Hamiltonian.

For the rotated expression, the off-diagonal term $\mathcal{N}_{xz}$ also appears and reads as
\begin{equation}
\begin{split}
    \mathcal{N}_{xz} \propto -\,(2S) \, \int d^2\vec{k}\,\,(2 \pi)^2 e^{-2kd} k_x^2\,\, \sin 2 \theta_m \\ \frac{1}{4} \,\left[\delta(\omega - \omega_{+,\vec{k}}) (u_{11} - u_{12} + v_{11}- v_{21})^2 \right.\\
    +\left. \delta(\omega - \omega_{-,\vec{k}}) (u_{12} - u_{22} + v_{12}- v_{22})^2  \right].
    \end{split}
\end{equation}

\subsection*{Lattice Fourier transformation of the dipole tensor}

In the scenario where $d \gg a$ does not hold and in particular for $d \ll a$, the approximation that we made previously by interpreting terms as $\left[ \sum_i D^{\alpha\gamma}(\vec{r}_i + \vec{\eta}) e^{-i\vec{k}\cdot (\vec{\rho}_i +\vec{\eta})}\right]$ as a continuum Fourier transform of $D(\vec{r})$, is not valid. For these cases, the sub-lattice information does not collapse and we need to keep it explicit.

For clarity, let us define

\begin{equation}
    \mathbb{D}^{\alpha \gamma}(\vec{k},\vec{\eta}) = \frac{1}{\sqrt{N}} \sum_i D^{\alpha\gamma}(\vec{r}_i + \vec{\eta}) e^{-i\vec{k}\cdot (\vec{\rho}_i +\vec{\eta})}
\end{equation}
and
\begin{equation}
    \mathbb{C}_{\gamma \delta}^{\eta \xi}(\vec{k},\omega)=-\Im (-i \int_0^\infty dt e^{i\omega t} \langle [S^\gamma_{\eta,\vec{k}}(t),S^\delta_{\xi,-\vec{k}}(0)] \rangle ).
\end{equation}
Hence, in direct analogy to what we derived before (for $d \gtrsim a$), we have

\begin{equation}
\begin{split}
    \mathcal{N}_{zz}  =\frac{1}{2} \coth{\frac{\omega}{2T}} \left(\frac{\mu_0 \mu_{\mathrm{B}}}{4 \pi}\right)^2  \int d^2\vec{k} \,\,
   \sum_{\eta, \xi} \sum_{\gamma,\delta} \left[\mathbb{D}^{z \gamma}(\vec{k},\vec{\eta}) \mathbb{D}^{z \delta}(-\vec{k},\vec{\xi})\mathbb{C}^{\eta \xi}_{\gamma \delta}(\vec{k},\omega)\right],
    \end{split}
\end{equation}
since the same symmetry constraints hold, we can write
\begin{equation}
\begin{split}
    \mathcal{N}_{zz}  =\frac{1}{2} \coth{\frac{\omega}{2T}} \left(\frac{\mu_0 \mu_{\mathrm{B}}}{4 \pi}\right)^2  \int d^2\vec{k} \,\,
   \sum_{\eta, \xi} &\left[  \mathbb{D}^{z x}(\vec{k},\vec{\eta}) \mathbb{D}^{z x}(-\vec{k},\vec{\xi})\mathbb{C}^{\eta \xi}_{xx}(\vec{k},\omega) +\mathbb{D}^{z y}(\vec{k},\vec{\eta}) \mathbb{D}^{z y}(-\vec{k},\vec{\xi})\mathbb{C}^{\eta \xi}_{yy}(\vec{k},\omega) \right.\\
   &\left. + \mathbb{D}^{z z}(\vec{k},\vec{\eta}) \mathbb{D}^{z z}(-\vec{k},\vec{\xi})\mathbb{C}^{\eta \xi}_{zz}(\vec{k},\omega)\right],
    \end{split}
\end{equation}

and 

\begin{equation}
    \begin{split}
        \mathbb{C}^{11}_{xx}(\vec{k},\omega)&= (2 S) \, \sin^2\theta_m  \frac{1}{4} \left[ (u_{11}+v_{11})^2\,\,\delta(\omega - \omega_{+,\vec{k}}) + (u_{12}+v_{12})^2\,\,\delta(\omega - \omega_{-,\vec{k}}) \right]
        \\
                \mathbb{C}^{12}_{xx}(\vec{k},\omega)+\mathbb{C}^{21}_{xx}(\vec{k},\omega)&=- (2 S) \, \sin^2\theta_m  \frac{2}{4}\left[(u_{11} + v_{11}) (u_{21} + v_{21}) \,\,\delta(\omega - \omega_{+,\vec{k}})\right.
                \\
              &  \left. + (u_{12} + v_{12}) (u_{22} + v_{22})\,\, \delta(\omega - \omega_{-,\vec{k}}) \right]
        \\
                \mathbb{C}^{22}_{xx}(\vec{k},\omega)&= (2 S) \, \sin^2\theta_m  \frac{1}{4}\left[(u_{21} + v_{21})^2 \,\,\delta(\omega - \omega_{+,\vec{k}}) + (u_{22} + v_{22})^2 \,\, \delta(\omega - \omega_{-,\vec{k}}) \right].
        \\
    \end{split}
\end{equation}
For the remaining correlators, one finds similar expressions.

\subsection{Relaxation rate}
\label{app:relaxation}
Similarly to the decoherence rate, we can relate the relaxation rate to a noise component. In this case, the transversal noise is the relevant one, e.g., for the qubit quantization axis along $\hat{z}$ we have $1/T_1 \propto \mathcal{N}_{-+}$ with these components being defined as $(\pm) = \hat{x} \pm i \hat{y}$, i.e., as written out explicitly in \equref{OneOverT1} below. If we now rotate into an arbitrary direction through $R_{\text{tot}} = R_x(\phi)R_z(\theta)$, we obtain

\begin{equation}
\begin{split}
\mathcal{N}_{-+} \xrightarrow{R_{tot}} &  \mathcal{N}_{-+} 
+ \sin^2\!\theta\left(\mathcal{N}_{zz} 
- \mathcal{N}_{xx}\sin^2\!\phi 
- \mathcal{N}_{yy}\cos^2\!\phi\right) \\
&+ 2i\left[\,\mathcal{N}_{xy}(\cos\theta - 1) 
+\,\mathcal{N}_{yz}\sin\theta\sin\phi\right]
- \mathcal{N}_{xz}\sin 2\theta\,\sin\phi
\end{split}
\end{equation}
 Here, we already omitted the vanishing terms proportional to $(\mathcal{N}_{xy} + \mathcal{N}_{yx}),(\mathcal{N}_{xz} - \mathcal{N}_{zx}) $ and $(\mathcal{N}_{yz} + \mathcal{N}_{zy})$, which are zero due to $\mathcal{N}_{xy}$, $\mathcal{N}_{yz}$, and $\mathcal{N}_{xz}$ being  purely imaginary and real, respectively.

\begin{figure}[tbh]
    \centering
    \includegraphics[width=0.5\linewidth]{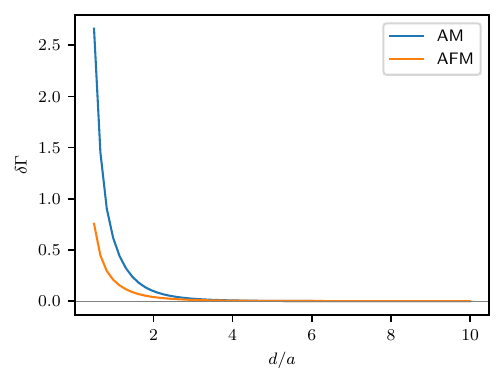}
    \caption{$\delta\Gamma$ for an AM ($\eta = 1/\sqrt{2}$) and an AFM ($\eta = 1$). Here the qubit axis is aligned along $\pm \hat{\vec{z}}$, $\omega = 0.25$, $h=0.5$ and $T=1$.}
    \label{fig:app_deltaGamma}
\end{figure}

The biggest difference with respect to the parallel noise $\mathcal{N}_{\parallel}$, is the presence of the fully imaginary $\mathcal{N}_{xy}$ and $\mathcal{N}_{yz}$. We remark, once more, that for the continuum approximation of $D^{\alpha \gamma}(\vec{k})$, they are zero. Therefore, for $d\gg a$, they vanish. However $d \ll a$, as shown before, requires the use of $\mathbb{D}^{\alpha \gamma} (\vec{k},\eta)$. In this case, we obtain
\begin{equation}
\begin{split}
    \mathcal{N}_{xy}  =\frac{1}{2} \coth{\frac{\omega}{2T}} \left(\frac{\mu_0 \mu_{\mathrm{B}}}{4 \pi}\right)^2  \int d^2\vec{k} \,\,
\sum_{\eta, \xi} \left[\mathbb{D}^{x x}(\vec{k},\vec{\eta}) \mathbb{D}^{yy}(-\vec{k},\vec{\xi}) - \mathbb{D}^{x y}(\vec{k},\vec{\eta}) \mathbb{D}^{yx}(-\vec{k},\vec{\xi}) \right] \mathbb{C}^{\eta \xi}_{xy},
\end{split}
\end{equation}
and a similar expression for $\mathcal{N}_{yz}$. Note that we used the fact that $\mathbb{C}^{\eta \xi}_{xy} = - \mathbb{C}^{\eta \xi}_{yx}$.

Finally, to connect to our approach for the itinerant electrons in the main text, we can calculate $\delta \Gamma_{\phi,\theta} = 1/T_1(\phi,\theta) - 1/T_1(\phi+\pi, \pi-\theta )$, e.g., $\delta \Gamma = \delta \Gamma_{\phi,0} = 1/T_1(\hat{\vec{z}}) - 1/T_1(-\hat{\vec{z}})$,
\begin{equation}
    \delta\Gamma(\phi,\theta) = 4i\,\mathcal{N}_{xy}\cos\theta + 4i\,\mathcal{N}_{yz}\sin\theta\sin\phi.
\end{equation}
In \figref{fig:app_deltaGamma} we can see that, for small distances, both AM and AFM show a finite values of $\delta \Gamma$. We have further verified that for vanishing Zeeman field $h=0$, $\delta \Gamma =0$ always, as expected by symmetry: for the AM, $C_{2z}\Theta$ leads to the vanishing of $\delta \Gamma$ while the reflection symmetry $\sigma_v$ does for the AFM, as is easily shown from \equsref{FirstNConstraint}{SecondNConstraint}.

\section{Transformation of spin-orbit coupling}\label{TransformationSOC}
Almost all of the terms we use for the effective semi-classical Hamiltonian around the domain wall have been derived in \refcite{2026arXiv260220236S}. The only exception is the effect of the slowly varying texture on the spin-orbit coupling term $\propto \alpha_R$ in \equref{EffectiveHamiltonian}. To discuss this, we first note that the unitary transformations used in this work obey
\begin{equation}
    \vec{A} := i U^\dagger \vec{\nabla} U \propto s_x,s_y, \quad \text{i.e.,} \quad P_- \vec{A} P_- = 0, \label{EmergentGaugeField}
\end{equation}
where $P_\pm = \frac{1}{2}(\mathbbm{1} \mp s_z \tau_z)$ projects onto the low- or high-energy subspace. This is the reason why there are no emergent gauge fields for the first two terms in \equref{EffectiveHamiltonian}. Intuitively, this is connected with the fact that the texture is effectively co-planar in the limit $R\rightarrow \infty$ (negligible curvature of the domain wall) that we focus on here. As such, there is no emergent magnetic field and a gauge exists where the associated emergent gauge field vanishes.

With this in mind, we now consider the projection of the rotated spin-orbit term, which is of the form (here $\hat{k}_j = - i \partial_j$ is the momentum operator)
\begin{align}
    \sum_{j,k=x,y}\alpha_{j,k} P_- U^\dagger \tau_x \hat{k}_j s_k U P_- &= \frac{1}{2}\sum_{j,k=x,y}\alpha_{j,k} P_- U^\dagger \tau_x \{ \hat{k}_j, s_k \} U P_- \\
    &= \frac{1}{2}\sum_{j,k=x,y}\alpha_{j,k} P_-  \tau_x \{ U^\dagger \hat{k}_j U, \tilde{s}_k \}  P_-, \quad \tilde{s}_k = U^\dagger s_k U \\
    &= \frac{1}{2}\sum_{j,k=x,y}\alpha_{j,k} P_-  \tau_x \{ \hat{k}_j - A_j , \tilde{s}_k \}  P_-, \,\, A_j = - U^\dagger (\hat{k}_j U) \\
    &= \frac{1}{2}\sum_{j,k=x,y}\alpha_{j,k} P_- \tau_x \left(    \{ \hat{k}_j , \tilde{s}_k^\perp  \}  - \{A_j ,  \tilde{s}_k^\parallel  \} \right) P_-,
\end{align}
where we suppressed the explicit dependence on $\vec{r}$ in $U$ and $\tilde{s}_k$ for brevity.
In the last line, we wrote $\tilde{s}_k = \tilde{s}_k^\parallel + \tilde{s}_k^\perp$ with $\tilde{s}_k^\parallel \propto s_z$ and $\tilde{s}_k^\perp \propto s_x,s_y$, or explicitly
\begin{equation}
    \tilde{s}_k^\perp = \frac{1}{2}\sum_{j=x,y} s_j \text{tr}[\tilde{s}_k s_j] = \frac{1}{2}\sum_{j=x,y} s_j \text{tr}[U^\dagger s_k U s_j]
\end{equation}
and $\tilde{s}_k^\parallel = \tilde{s}_k - \tilde{s}_k^\perp$. We further used \equref{EmergentGaugeField}. The latter in turn also implies $\{A_j ,  \tilde{s}_k^\parallel  \} =0$. In total and in the semi-classical limit (see, e.g., \cite{2026arXiv260220236S}), we therefore get a spin-orbit term of the form
\begin{equation}
    \alpha_R \left( k_y \widetilde{\sigma}_x(\vec{r}) - k_x \widetilde{\sigma}_y(\vec{r}) \right), \quad \widetilde{\sigma}_j(\vec{r}) = \frac{1}{2}\sum_{j'=x,y} \sigma_{j'}\text{tr}[U^\dagger(\vec{r}) \sigma_j U(\vec{r}) \sigma_{j'}],
\end{equation}
after transformation and projection.

\section{Symmetry constraints on noise}\label{SymmetriesOnN}

\subsection{General remarks}
As already discussed in detail in the main text, the central object that NV noise magnetometry is sensitive to is the correlator 
\begin{equation}
    \mathcal{N}_{ab}(\Omega) = \frac{1}{2}\int_{-\infty}^\infty \diff t \, e^{i\Omega t} \braket{\{ B_a(t), B_b(0) \}} \label{Definition}
\end{equation}
of the time-dependent magnetic field $\vec{B}(t)$ at the location of the NV center. We begin by discussing some symmetry constraints on it. First, let us assume that the system is \textit{stationary}. Then
\begin{equation}
    \mathcal{N}_{ab}(\Omega) = \frac{1}{2}\int_{-\infty}^\infty \diff t \, e^{i\Omega t} \braket{\{ B_a(0), B_b(-t) \}} = \frac{1}{2}\int_{-\infty}^\infty \diff t \, e^{-i\Omega t} \braket{\{ B_b(t), B_a(0) \}} = \mathcal{N}_{ba}(-\Omega). \label{Stationary}
\end{equation}
Complex conjugation of the definition (\ref{Definition}) leads to $\mathcal{N}^*_{ab}(\Omega) = \mathcal{N}_{ab}(-\Omega)$ and, thus, the \textit{reality condition}
\begin{equation}
    \mathcal{N}^\dagger(\Omega) = \mathcal{N}(\Omega). \label{DaggerConditionOnN}
\end{equation}
Finally, under \textit{time-reversal symmetry}, it holds
\begin{equation}
    \mathcal{N}_{ab}(\Omega) \rightarrow \frac{1}{2}\int_{-\infty}^\infty \diff t \, e^{i\Omega t} \braket{\{-B_a(-t), - B_b(0) \}} = \frac{1}{2}\int_{-\infty}^\infty \diff t \, e^{-i\Omega t} \braket{\{B_a(t),B_b(0) \}} = \mathcal{N}_{ab}^*(\Omega) = \mathcal{N}_{ba}(\Omega) , \label{TRSConstraint}
\end{equation}
where we used \equref{DaggerConditionOnN} in the last step; this makes the action of time-reversal symmetry very natural as it just corresponds to transposition akin to Onsager relations. 

To understand the relation to the symmetries in the material that is probed, we have to take into account the relation between $\vec{B}(t)$ and the material properties, which are coupled through \equref{eq:B} or \equref{BiotSavart}, depending on the type of fluctuations. Let us assume that the material is invariant under the symmetry transformation $g$ which is not broken (at least approximately) by the position of the NV center (e.g., elementary lattice translations are approximately preserved for $d \gg a$). We then have
\begin{equation}
    \mathcal{N}(\Omega) = R^T(g)\mathcal{N}(\Omega) R(g), \label{FirstNConstraint}
\end{equation}
where $R(g)$ is the appropriate pseudo-vector representation of $g$. Similarly, if the system exhibits a magnetic symmetry, i.e., $g \Theta$ involving time-reversal $\Theta$, we have
\begin{equation}
    \mathcal{N}(\Omega) = R^T(g)\mathcal{N}^T(\Omega) R(g). \label{SecondNConstraint}
\end{equation}

Denoting the qubit axis by $\hat{\vec{n}}$, the relaxation rates can be obtained as follows:
\begin{equation}
    \frac{1}{T_2(\hat{\vec{n}})} \propto \int \diff \Omega \, f(\Omega) \, \hat{\vec{n}}^T\mathcal{N} (\Omega)\hat{\vec{n}},
\end{equation}
where $f$ is some protocol-dependent filter function, that does not affect our symmetry analysis here. Furthermore, it holds
\begin{equation}
    \frac{1}{T_1(\hat{\vec{n}}_{\text{q}})} \propto \widetilde{\mathcal{N}}_{xx}(\omega_{\text{q}}) + \widetilde{\mathcal{N}}_{yy}(\omega_{\text{q}}) + 2\,\text{Im}[\widetilde{\mathcal{N}}_{xy}(\omega_{\text{q}})], \label{OneOverT1}
\end{equation}
where $\omega_{\text{q}}$ is the qubit frequency and $\widetilde{\mathcal{N}} = R_{\hat{\vec{n}}}\mathcal{N} R^T_{\hat{\vec{n}}}$ with $R_{\hat{\vec{n}}}$ rotating $\hat{\vec{z}}$ onto $\hat{\vec{n}}$.

With this at hand, it is straightforward to derive the symmetry constraints on the components of $\mathcal{N}$ and the resulting features of the relaxation rates.
For instance, we get $\delta \Gamma \propto \text{Im}[\mathcal{N}_{xy}(\omega_{\text{q}}) - \mathcal{N}_{yx}(\omega_{\text{q}})] = 2 \text{Im}[\mathcal{N}_{xy}(\omega_{\text{q}})]$ as stated in the main text. We now have, for instance, 
\begin{equation}
    R(\sigma_d) = \begin{pmatrix} 0 & -1 & 0 \\ -1 & 0 & 0 \\ 0 & 0 & -1 \end{pmatrix}, \quad \text{such that} \quad \sigma_d: \,\, \mathcal{N}_{xy}(\omega_{\text{q}}) \rightarrow \mathcal{N}_{yx} (\omega_{\text{q}}) = \mathcal{N}^*_{xy} (\omega_{\text{q}}).
\end{equation}
Consequently, the presence of $\sigma_d$ implies that $\delta \Gamma = 0$. The same holds for $C_{4z}\Theta$.

\subsection{Expressions for the angular dependence}

In analogy to what we did in \secref{MagneticNoise} and \appref{app:noise}, we define the rotation $R_{\text{tot}} = R_x(\phi)R_z(\theta)$ and the rotated expressions for the relaxation and decoherence rates. Note that we are using the symmetry constraints discussed in \secref{SymmetryConsiderations} for AFMs and AMs with itinerant electrons. Explicitly, the expressions for the four combinations of $T_1^{-1}$, $T_2^{-1}$ and AFM, AM read as

\begin{align}
T_2^{-1}\,(\text{AFM}) ={}& \sin^2(\theta)\left(\mathcal{N}_{xx}\sin^2(\phi) + \mathcal{N}_{yy}\cos^2(\phi)\right) + \mathcal{N}_{zz}\cos^2(\theta), \\[1em]
T_1^{-1}\,(\text{AFM}) ={}& 2i\mathcal{N}^B_{xy}\cos(\theta) + \sin^2(\phi)\left(\mathcal{N}_{xx}\cos^2(\theta) + \mathcal{N}_{yy}\right) \nonumber \\
&+ \mathcal{N}_{xx}\cos^2(\phi) + \mathcal{N}_{yy}\cos^2(\theta)\cos^2(\phi) + \mathcal{N}_{zz}\sin^2(\theta), \\[1em]
T_2^{-1}\,(\text{AM})  ={}& -2\mathcal{N}^B_{yz}\sin(\theta)\cos(\theta)\cos(\phi) + \sin^2(\theta)\left(\mathcal{N}_{xx}\sin^2(\phi) + \mathcal{N}_{yy}\cos^2(\phi)\right) + \mathcal{N}_{zz}\cos^2(\theta), \\[1em]
T_1^{-1}\,(\text{AM})  ={}& 2i\mathcal{N}^B_{xy}\cos(\theta) + 2\sin(\theta)\cos(\phi)\left(\mathcal{N}^B_{yz}\cos(\theta) + i\mathcal{N}_{xz}\right) \nonumber \\
&+ \cos^2(\theta)\left(\mathcal{N}_{xx}\sin^2(\phi) + \mathcal{N}_{yy}\cos^2(\phi)\right) + \mathcal{N}_{xx}\cos^2(\phi) + \mathcal{N}_{yy}\sin^2(\phi) + \mathcal{N}_{zz}\sin^2(\theta),
\end{align}

where $\mathcal{N}_{xy}^B$ and $\mathcal{N}_{xz}^B$ are purely imaginary.

\end{appendix}

\end{document}